\documentclass[12pt]{article}
\usepackage{epsf,amsfonts,hyperref}
\renewcommand{\appendix}[1]{
    \setcounter{equation}{0}
    \renewcommand{\thesection}{\Alph{section}}
    \section{Appendix: \protect\indent #1}
}
\newcommand\encadremath[1]{\vbox{\hrule\hbox{\vrule\kern8pt
\vbox{\kern8pt \hbox{$\displaystyle #1$}\kern8pt}
\kern8pt\vrule}\hrule}}
\def\enca#1{\vbox{\hrule\hbox{
\vrule\kern8pt\vbox{\kern8pt \hbox{$\displaystyle #1$}
\kern8pt} \kern8pt\vrule}\hrule}}

\newcommand\figureframex[3]{
\begin{figure}[bth]
\hrule\hbox{\vrule\kern8pt
\vbox{\kern8pt \vbox{
\begin{center}
{\mbox{\epsfxsize=#1.truecm\epsfbox{#2}}}
\end{center}
\caption{#3}
}\kern8pt}
\kern8pt\vrule}\hrule
\end{figure}
}
\newcommand\figureframey[3]{
\begin{figure}[bth]
\hrule\hbox{\vrule\kern8pt
\vbox{\kern8pt \vbox{
\begin{center}
{\mbox{\epsfysize=#1.truecm\epsfbox{#2}}}
\end{center}
\caption{#3}
}\kern8pt}
\kern8pt\vrule}\hrule
\end{figure}
}

\renewcommand{\thesection}{\arabic{section}}

\makeatletter
\@addtoreset{equation}{section}
\makeatother
\newtheorem{theorem}{Theorem}[section]

\newtheorem{remark}{Remark}[section]
\newtheorem{proposition}{Proposition}[section]
\newtheorem{lemma}{Lemma}[section]
\newtheorem{corollary}{Corollary}[section]
\newtheorem{definition}{Definition}[section]
\def\br{\begin{remark}\rm\small}
\def\er{\end{remark}}
\def\bt{\begin{theorem}}
\def\et{\end{theorem}}
\def\bd{\begin{definition}}
\def\ed{\end{definition}}
\def\bp{\begin{proposition}}
\def\ep{\end{proposition}}
\def\bl{\begin{lemma}}
\def\el{\end{lemma}}
\def\bc{\begin{corollary}}
\def\ec{\end{corollary}}
\def\beaq{\begin{eqnarray}}
\def\eeaq{\end{eqnarray}}
\newcommand{\proof}[1]{{\noindent \bf proof:}\par
{#1} $\square$}

\newcommand{\eq}[1]{eq.(\ref{#1})}

\newcommand{\beq}{\begin{equation}}
\newcommand{\eeq}{\end{equation}}
\newcommand{\bea}{\begin{eqnarray}}
\newcommand{\eea}{\end{eqnarray}}

\renewcommand{\and}{{\qquad {\rm and} \qquad}}

\newcommand{\virg}{{\qquad , \qquad}}

 \newcommand{\Tr}{{\,\rm Tr}\:}
\newcommand{\tr}{{\,\rm tr}\:}

\newcommand{\Res}{\mathop{\,\rm Res\,}}

\newcommand{\td}[1]{{\tilde{#1}}}

\renewcommand{\l}{\lambda}
\newcommand{\om}{\omega}

\newcommand{\ee}[1]{{{\rm e}^{#1}}}

\newcommand{\Pint}{{\int\kern -1.em -\kern-.25em}}

\renewcommand{\l}{\lambda}
\renewcommand{\L}{\Lambda}

\newcommand{\ovl}{\overline}

\begin{document}
%=============================Page de titre==============%\date{??}
%\author{Eynard}
%\title{Correlation functions for hermitian random matrices}
%\topmargin .5cm \textheight 21.5cm \textwidth 15.8cm
%\oddsidemargin 0.54cm
%\evensidemargin 0.54cm
\sloppy

%\maketitle

\pagestyle{empty}
\hfill IPhT-T09/116
\addtolength{\baselineskip}{0.20\baselineskip}
\begin{center}
\vspace{26pt}
{\large \bf {Universal scaling limits of matrix models,\\ and $(p,q)$ Liouville gravity}}
\newline
\vspace{26pt}

{\sl M.\ Berg\`ere}\hspace*{0.05cm}\footnote{ E-mail: michel.bergere@cea.fr },
{\sl B.\ Eynard}\hspace*{0.05cm}\footnote{ E-mail: bertrand.eynard@cea.fr },
\vspace{6pt}
Institut de Physique Th\'eorique,\\
CEA, IPhT, F-91191 Gif-sur-Yvette, France,\\
CNRS, URA 2306, F-91191 Gif-sur-Yvette, France.\\
\end{center}

\vspace{20pt}
\begin{center}
{\bf Abstract}:

We show that near a point where the equilibrium density of eigenvalues of a matrix model behaves like $y\sim x^{p/q}$, the correlation functions of a random matrix, are, to leading order in the appropriate scaling, given by determinants of the universal $(p,q)$-minimal models kernels.
Those $(p,q)$ kernels are written in terms of functions solutions of a linear equation of order $q$, with polynomial coefficients of degree $\leq p$. For example, near a regular edge $y\sim x^{1/2}$, the $(1,2)$ kernel is the Airy kernel.
Those kernels are associated to the $(p,q)$ minimal model, i.e. the $(p,q)$ reduction of the KP hierarchy solution of the string equation.
Here we consider only the 1-matrix model, for which $q=2$.

\end{center}

%-----------------------------ABSTRACT--------------------------------------
%
%Abstract

%\begin{center}

%\end{center}

%\newpage
%\pagestyle{empty}

%\section*{}

%\newpage
\vspace{26pt}
\pagestyle{plain}
\setcounter{page}{1}

%*********************************************************************
%==================== ARTICLE =======================================%*********************************************************************

\tableofcontents

\section{Introduction}

In this article, we shall consider "scaling limits" of matrix integrals.

We shall show, under certain assumptions, that scaling limits of matrix integrals are governed by some well known integrable systems.
The fact that double scaling limits of matrix models are minimal models $(p,q)$ of conformal field theories \cite{BookPDF}, has been well known in the physics literature for a long time (see \cite{DGZ, Cicuta, gross:1991} for review, and among others see \cite{KPZ, KazakovRMTcrit, Kazakovloop, KazakovRMT, Moo, Kos, BKa, DS, GM, DKK, GM}), and here we merely summarize some results scattered in the physics literature, we present the main features of those universal limit laws, and provide a mathematical proof.

The idea of the proof works backwards: we show that $(p,q)$ minimal models determinantal correlation functions satisfy the same recursion as the scaling limits of matrix models.

\medskip
We shall consider only the 1-matrix model, whose corresponding limit integrable systems are the $(p,2)$ minimal models, reductions of KdV, and we hope to later generalize those results to multi-matrix models and general $(p,q)$ limits, as claimed in many physics works \cite{DGZ, Kazakovloop, KPZ}.

\medskip
The main result, theorem \ref{mainth}, is that limit correlation functions are given by determinantal formulae of the $(p,q)$ kernel.

\subsubsection{Example: $(1,2)$ law: Airy kernel and Tracy-Widom law}

The equilibrium density of eigenvalues of a $N\times N$ random hermitian matrix, generically behaves near the edge of the distribution, like:
\beq
\rho(x) \sim x^{1\over 2}.
\eeq
It is well known that, after rescaling $x$ by $N^{2/3}$, the $n-$points correlation functions in the vicinity of the edge, are given by determinants of the Airy kernel which appears in Tracy-Widom law \cite{TW} of extreme eigenvalues statistics:
\beq
\rho_n(N^{-2/3}x_1,\dots,N^{-2/3}x_n) \mathop{{\sim}}_{N\to\infty} N^{{2n\over 3}}\det(\hat K_{\rm Airy}(x_i,x_j))\,\, (1+O(N^{-1/3})),
\eeq
and the Airy kernel is the Christoffel-Darboux kernel of the Airy function:
\beq
\hat K_{\rm Airy}(x_1,x_2)  = {Ai(x_1)Ai'(x_2)-Ai'(x_1)Ai(x_2)\over x_1-x_2}.
\eeq
Notice that the Airy function satisfies a 2nd order ODE, whose coefficients are polynomials of degree $1$:
\beq
Ai''(x) =x\, Ai(x),
\eeq
which can also be written as a $2\times 2$ differential system:
\beq
{d\over dx}\,\, \pmatrix{Ai(x)\cr Ai'(x)} = \pmatrix{ 0 & 1 \cr x & 0}\,\pmatrix{Ai(x)\cr Ai'(x)}.
\eeq

\subsubsection{Higher $(p,2)$ laws}

\figureframex{14}{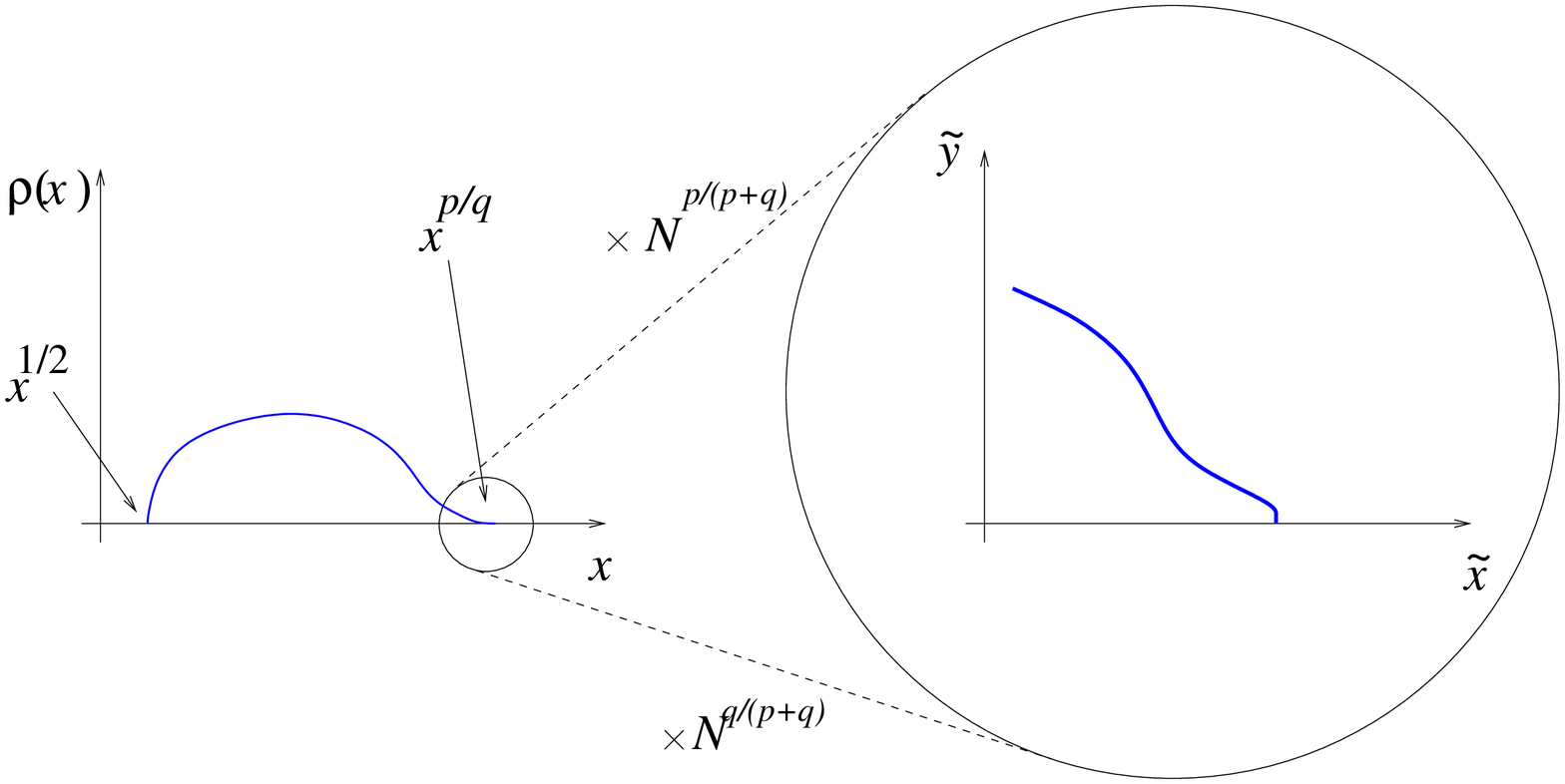}
{Generically, the equilibrium density of eigenvalues behaves like $x^{1/2}$ near endpoints. It may happen, after fine tuning the parameters, that it behaves like $x^{p/q}$ ($q=2$ for 1-matrix model). The eigenvalue statistics at a scale $\td x=x\,\,N^{q/(p+q)}$ is governed by the universal $(p,q)$ kernel law. The $(p,q)$ kernel is an integrable kernel, associated to the $(p,q)$ reduction of the KP hierarchy. For a regular endpoint $(p,q)=(1,2)$ it is the  Airy kernel.}

More generally consider a $p/q$ singularity of the equilibrium density of eigenvalues:
\beq
\rho(x) \sim x^{p/q}.
\eeq
We shall consider only $q=2$ and $p=2m+1$ in this article, but we recall that the physics literature claims that general $y\sim x^{p/q}$ case can be treated the same way, and should correspond to $(p,q)$ minimal models.

We shall see, after rescaling $x$ by $N^{q/(p+q)}$, 
that the correlation functions in the vicinity of the edge, are given by determinants of the $(p,q)$ kernel which appears in $(p,q)$ minimal models of conformal field theory \cite{BookPDF}.
\beq
\rho_n(N^{-{q\over p+q}}x_1,\dots,N^{-{q\over p+q}} x_n) \mathop{{\sim}}_{N\to\infty} 
N^{n\,{q\over p+q}} \det(\hat K_{(p,q)}(x_i,x_j))\,\, (1+O(N^{-1/(p+q)})),
\eeq
and the $(p,2)$ kernel is the Christoffel-Darboux kernel of the $(p,2)$ Baker-Akhiezer function:
\beq
\hat K_{(p,2)}(x_1,x_2)  = {\psi(x_1)\td\psi(x_2)-\td\psi(x_1)\psi(x_2)\over x_1-x_2}.
\eeq
where the $(\psi,\td\psi)$ functions satisfy a 2nd order ODE:
\beq
{d\over dx}\,\, \pmatrix{\psi(x)\cr\td\psi(x)} = {\cal D}_{(p,q)}(x)\,\pmatrix{\psi(x)\cr\td\psi(x)}
\eeq
where ${\cal D}_{(p,2)}(x)$ is a $2\times 2$ matrix with polynomial coefficients, such that the degree of $\det {\cal D}$ is of degree at most $p$.

Moreover this differential system, is associated to the Lax matrix of the $(p,2)$ reduction of the integrable KdV hierarchy, which means that the coefficients of the matrix ${\cal D}_{(p,q)}(x)$, are themselves solution of some non-linear integrable differential equations.

The coefficients of ${\cal D}_{(p,q)}(x)$ are differential polynomials of a function $u(t)$ which satisfies (for $p=2m+1$, $q=2$), the $m+1^{\rm th}$ Gelfand-Dikii non linear equation \cite{DGZ}, of the form:
\beq
u^{m+1} + u^{m-1} \ddot u + \dots + u^{(2m)} = t.
\eeq
The time $t$ measures the distance to the critical point.

For example for pure gravity $(p,q)=(3,2)$, ${\cal D}_{(3,2)}(x)$ is a $2\times 2$ matrix, with polynomial coefficients such that the degree of $\det {\cal D}_{(3,2)}(x)$ is $3$:
\beq
{\cal D}_{(3,2)}(x,t) = \pmatrix{{\dot u} & 2x-2u  \cr (x+2u)(2x-2u)+{\ddot u} & -\, {\dot u}}
\eeq
and where $u(t)$ is solution of the Painlev\'e I equation ($1^{\rm st}$ Gelfand-Dikii equation):
\beq
3 u^2(t) - {\ddot u(t)\over 2}=t.
\eeq

All this has been stated for a long time in the physics literature, and we shall just present it concisely and prove it.

\subsection{Universality of eigenvalues statistics point of view}
\label{secuniversality}

In this subsection, we summarize some well known facts about random matrices \cite{Mehtabook, Wigner, Dysondet,  BIbook, Moerbeke:2000, DGZ, BPS, BZ, Joh}, and we fix the notations.

\medskip
Consider a probability law of the form of the joint law of eigenvalues of a random hemitian type\footnote{Hermitian matrices correspond to real eigenvalues and positive measure $d\mu$, but it is customary to generalize random matrices to normal matrices having their eigenvalues on some contours in the complex plane, and the measure $d\mu$ can be complex. The loop equations are the same for all those models, they are independent of the integration contour, and thus, they can all be treated in the same framework.} matrix:
\beq
d\mu(\l_1,\dots,\l_N) = {1\over Z} \, \prod_{i<j}(\l_j-\l_i)^2\,\, \prod_i \ee{-{N\over s} V(\l_i)}\,\,d\l_i  ,
\eeq
where $Z$ is the partition function
\beq
Z = \int \, d\mu = \int dM\,\, \ee{-{N\over s}\tr V(M)}.
\eeq
Here $s$ is a parameter, often called the {\em temperature}.
we shall be interested in the large $N$ limit, and possibly a limit  $s\to s_c$, where $Z$ has a singularity at $s=s_c$.
The name {\bf double scaling limit} \cite{Kazakovloop, DGZ, KazakovRMTcrit, GM, gross:1991, DSS, DS, DKK, BI2} means that we consider a regime where the limits $s\to s_c$ and $N\to \infty$ are related by a scaling relation
\beq
\encadremath{
(s-s_c)\,N^{-\alpha} = O(1)
}\eeq
where $\alpha$ is some appropriate exponent ($\alpha=0$ if $Z$ is not singular).

We are interested in computing expectation values of resolvents:
\beq
\bar {\hat \om}_n(x_1,\dots,x_n) = \left< \Tr{1\over x_1-M} \dots \Tr {1\over x_n-M}\right> 
= \left< \sum_{i_1,\dots,i_n} {1\over x_1-\l_{i_1}}\dots {1\over x_n-\l_{i_n}} \right>
\eeq
as well as in their cumulants
\beq
\hat\om_n(x_1,\dots,x_n) = \left< \Tr{1\over x_1-M} \dots \Tr {1\over x_n-M}\right>_c .
\eeq
The density correlation functions $\rho_n(x_1,\dots,x_n)$ can be easily deduced from them: densities are discontinuities of resolvents, and resolvents are Stieljes transforms of densities, for example for the 1-point function:
\beq
\hat\om_1(x) = \int {\rho_1(x')dx'\over x-x'}
\virg
\rho_1(x) = {1\over 2i\pi}\,(\hat\om_1(x-i0)-\hat\om_1(x+i0)).
\eeq

\bigskip
Imagine, that, for $s<s_c$, the potential $V(x)$ is such that there is a large $N$ expansion of the type:
\beq
\ln Z = \sum_{g=0}^\infty (N/s)^{2-2g}\, \hat f_g,
\eeq
and similarly:
\beq
\hat\om_n(x_1,\dots,x_n) = \sum_{g=0}^{\infty} (N/s)^{2-2g-n}\, \hat\om_n^{(g)}(x_1,\dots,x_n).
\eeq
First, let us emphasize that such an expansion does not exist for any potential $V$, it exists only if the integration contour for the $\l_i$'s is a "steepest descent contour" for the potential $V$ (i.e. a landpath and bridges path in the Riemann-Hilbert language of \cite{Bertola, DKMVZ}).
For instance it was proved \cite{guionnet2} that such a large $N$ expansion holds for $s$ sufficiently small.

\smallskip
From now on, let us assume that we are in a situation where such an expansion exists.
In that case, the coefficients $\hat\om_n^{(g)}$ and $\hat f_g$ were computed in \cite{eynloop1mat, CEO, EOFg}, and they are the "spectral invariants" of some spectral curve associated to $V/s$.
The spectral curve $\hat y(x)$, in that case, is the function $\hat y(x)=V'(x)/2 - \hat \om_1^{(0)}(x)$, it is the large $N$ density, also called equilibrium density $\hat y(x)=i\pi\rho_{\rm eq}(x)=i\pi\rho_1^{(0)}(x)$:
\bt (proved in \cite{eynloop1mat, ec1loopF})\label{thWngsympinv}
The coefficents $\hat f_g$ and $\hat \om_n^{(g)}$ of the topological expansion of $\ln Z$ and $\hat \om_n$, are the spectral invariants (in the sense of \cite{EOFg}) of the spectral curve:
\beq
\hat y(x) = i\pi\, \rho_{\rm eq}(x) = {1\over 2}\,V'(x)-\hat \om_1^{(0)}(x).
\eeq
\et

\br  We refer the reader to \cite{EOFg} to see how to compute the spectral invariants of an arbitrary plane curve $\hat y(x)$. 
We shall give an explicit example of computation of  spectral invariants for formal matrix models below in section \ref{secspinv1}, see theorem \ref{thomngmmformalspinv}.

Let us say that we shall not be really using any deep result of \cite{EOFg} in this article, except the theorem 8.1. of \cite{EOFg} (which is very easy to prove by recursion).

\er

\medskip
Here, as is well known in random matrix theory, $\hat \om_1^{(0)}(x)$ is an algebraic curve (hyperelliptical for the 1-matrix model), with typical square-root branchpoints at the endpoints of the distribution of eigenvalues, we shall write it:
\beq
\hat \om_1^{(0)}(x) = {1\over 2} V'(x) - \hat y(x)
\eeq
where $\hat y(x)$ is the square root of some polynomial
\beq
\hat y^2 = {\rm Polynomial}(x).
\eeq

Generically, this $s$-dependent polynomial has only simple zeroes, and $\hat y(x)$ has square root singularities, but for some appropriate choices of $s=s_c$, the polynomial may have multiple zeroes, and we shall consider that, at $s=s_c$, there is a zero of order $2m+1$ at $x=0$:
\beq
s=s_c\,  \,\,\, \longrightarrow \qquad \hat y(x) \sim x^{m+{1\over 2}}.
\eeq
When $s$ is close to $s_c$, we typically have:
\beq
\hat y \sim \sum_{k=0}^m x^{k+{1\over 2}}\, c_k\,(s-s_c)^{{m-k\over m+1}} (1+O((s-s_c)^{1\over m+1})\,\,)\,\,,
\eeq
which we write:
\beq\label{eqdefspcurvelim}
\hat y((s-s_c)^{1\over m+1}\, x) \sim (s-s_c)^{2m+1\over 2m+2}\,\,y(x)\,\, (1+O((s-s_c)^{1\over m+1})\,\,
\virg
y(x) = \sum_{k=0}^m c_k\, x^{k+{1\over 2}}.
\eeq
At $s=s_c$ we have $\hat y\sim x^{m+{1\over 2}}$ and at $s\neq s_c$ we have $\hat y\sim \sqrt{x}$.

Notice that a regular endpoint corresponds to $m=0$, and in that case, $s_c$ can be chosen as any value of $s$.

\subsubsection{Double scaling limit}

In this article, we shall be interested in the behavior of $\hat \om_n^{(g)}$ when $s$ is close to $s_c$, and the $x_i$'s are in the vicinity of a branchpoint. 
Theorem 8.1. of  \cite{EOFg}, implies that after rescaling, we have (when $2-2g-n<0$):
\bt (theorem 8.1. of \cite{EOFg}):\label{thWnglimMM}
If $m>0$:
\beq
\hat f_g \sim (s-s_c)^{(2-2g){2m+3\over 2m+2}}\, f_g,
\eeq
and if $m\geq 0$:
\bea
\hat \om_n^{(g)}((s-s_c)^{1\over m+1}x_1,\dots,(s-s_c)^{1\over m+1}x_n) \cr
 \sim (s-s_c)^{(2-2g-n){2m+3\over 2m+2}-{n\over m+1}}\,\om^{(g)}_n(x_1,\dots,x_n)\,\,\, (1+O((s-s_c)^{{1\over m+1}} )),
\eea
where the $f_g$'s and  $\om_n^{(g)}$'s are the spectral invariants of \cite{EOFg} for the spectral curve $y(x)$ appearing in \eq{eqdefspcurvelim}.
\et
\br This theorem is very easy to prove by recursion on $n$ and $g$ from the definitions of spectral invariants in \cite{EOFg}. A more detailed proof is also given in section 4.8.2 of \cite{EOFgreview}.
\er

\medskip

Our goal in this article, is to show that the coefficients $\om_n^{(g)}$ and $f_g$ can also be computed from determinantal formulae for the $(p,2)$ kernel appearing in the $(p,2)$ minimal model, this is our theorem \ref{mainth}.

\subsubsection{Specific heat}

If $m>0$, we consider the resummation to leading order:
\beq
\ln Z = \sum_g (N/s)^{2-2g}\,\hat f_g \sim  \sum_{g} (N/s_c)^{(2-2g)} (s-s_c)^{(2-2g){2m+3\over 2m+2}}\, f_g = F((s-s_c)N^{2m+2\over 2m+3}).
\eeq
This shows that the double scaling limit is $N\to\infty$, $s\to s_c$ with a scaling:
\beq
\encadremath{
t = (s-s_c) \, N^{{2m+2\over 2m+3}} = O(1).
}\eeq 
This is a special case of the double scaling limit $(s-s_c)\sim N^{-(p+q-1)/(p+q)}$ for general $(p,q)$.

We defined the function
\beq\label{eqgenfgu}
F(t) = \sum_{g=0}^\infty t^{(2-2g){2m+3\over 2m+2}} f_g .
\eeq
Consider its second derivative, often called specific heat:
\beq
u(t) = {d^2\over dt^2}\, F(t) \,\, = \sum_g u_g\,t^{1-g(2m+3)\over m+1}.
\eeq
i.e.
\beq
u_g = (1-g)\,f_g\,\,{(2m+3)(m+2-g(2m+3))\over (m+1)^2}.
\eeq

We shall prove in this article, that, as claimed in many physics articles, this function satisfies the $m+1^{\rm th}$ Gelfand-Dikii non-linear equation.
For instance if $m=1$, it satisfies the Painlev\'e I equation:
\beq
3u^2 - {1\over 2} \ddot u = t.
\eeq

Moreover, it is well known from general considerations in statistical physics, that the free energy $-\ln Z$ should be convex, i.e. $u(t)$ should be negative for $t>0$:
\beq
u(t)\leq 0.
\eeq

\medskip
\br
The case $m=0$, needs some care. The correlation functions $\om_n$'s, indeed correspond to the $s\to s_c$ limits of $\hat\om_n$'s, i.e. a zoom $x\to (s-s_c) x$ near a regular branch point, but the free energy $F=\ln Z$ is not divergent at $s=s_c$, and thus $F(t)$ cannot be seen as the $s\to s_c$ limit of $\ln Z$.
In that case $m=0$, the $1$st Gelfand-Dikii equation is not differential, it is simply
\beq
u(t)=-{t\over 2},
\eeq
and, if it made sense, it would correspond to a free energy diverging as $F\sim N^2$, but in fact the free energy $F$ is not divergent, and one finds that $f_g=0$ for $g\geq 1$.
\er 

\subsection{Formal matrix models and combinatorics point of view}

As it was discovered by Brezin-Itzykson-Parisi-Zuber \cite{BIPZ}, matrix integrals are (in the formal sense) generating functions for counting discrete surfaces of a given topology \cite{davidRMT,PDFgraph, DGZ}.

Consider the potential:
\beq
V(x) = {x^2\over 2s} - {1\over s} \delta V(x)
\virg
\delta V(x) = \sum_{j=3}^{d+1} {s_j\over j}\, x^j
\eeq
Formal matrix integrals are defined as:
\beq\label{defformalint}
Z = \int_{\rm formal} dM\, \ee{-{N\over s}\Tr V(M)} \,\, \stackrel{{\rm def}}{=}\,\,  \sum_{k=0}^\infty\, {N^k\over k!\, s^k} \int_{H_N} dM\,\, (\tr \delta V(M))^k\,\, \ee{-{N\over s}\Tr {M^2\over 2}}.
\eeq
where $dM$ is the Lebesgue measure on $H_N$, normalized such that $\int dM\,\ee{-{N\over s}\Tr {M^2\over 2}} = 1$. In other words, we have exchanged the integral and the summation over $k$.
$Z$ is a formal power series in $s$, where each coefficient is a finite sum of polynomial expectation values of a Gaussian integral:
\beq
Z = 1+ \sum_{j=1}^\infty s^j \td A_j.
\eeq
After taking the Log, we also have a formal power series:
\beq
\ln Z = \sum_{j=1}^\infty s^j A_j.
\eeq
It was noticed by t'Hooft \cite{thooft} and then BIPZ\cite{BIPZ}, that, after dividing by $N^2$, each coefficient $A_j$ is a polynomial in $1/N^2$, namely:
\beq
A_j = \sum_{g=0}^{g_{\rm max}(j)}\,\, N^{2-2g}\,\, A_{j,g}.
\eeq
Then, one defines:
\beq
\hat f_g = \sum_{j=1}^\infty s^{j+2-2g}\,\, A_{j,g}
\eeq
which is also a formal series in powers of $s$ (one can easily prove that $A_{j,g}=0$ if $j+2-2g<0$).
In that case, the following large $N$ topological expansion holds as an equality between formal series of $s$, order by order in $s$:
\beq
\encadremath{
\ln Z = \sum_{g=0}^\infty (N/s)^{2-2g}\,\, \hat f_g.
}\eeq
We emphasize that this equality is not a large $N$ asymptotic expansion, it is a small $s$ asymptotic expansion, and order by order in $s$ the sum over $g$ is finite.

\medskip
It was proved by BIPZ in 1978 \cite{BIPZ}, by a mere application of Wick's theorem, that $\hat f_g$ is the generating function for maps of genus $g$:
\beq\label{eqfgnbmaps}
\hat f_g = \sum_{{\rm maps,\, genus}\,g} {s^{\#{\rm vertices}}\over \#{\rm Aut}}\,\,\, s_3^{\# {\rm triangles}}\, s_4^{\# {\rm quadrangles}} \dots \, s_{d+1}^{\# (d+1){\rm -gons}}.
\eeq
Similarly one may compute formal expectation values:
\bea
\hat \om_n(x_1,\dots,x_n) 
&=& \sum_{l_1,\dots,l_n}\, {1\over x_1^{l_1+1}\dots x_n^{l_n+1}}\,\left< \tr M^{l_1}\dots \tr M^{l_n} \right>_{c,{\rm formal}} \cr
&=& \left< \tr {1\over x_1-M}\dots \tr {1\over x_n-M} \right>_{c,{\rm formal}}
\eea
where $c$ means cumulant, and {\em formal} means that we compute the integral by exchanging the order of the Taylor expansion of $\ee{-{N\over s}\tr \delta V(M)}$ and the integral, as in \eq{defformalint}.
$\hat\om_n(x_1,\dots,x_n)$ is thus a formal power series in $s$, whose coefficients are polynomial expectation values of a Gaussian integral. For each power of $s$, the coefficient in $\hat\om_n$ is a polynomial in the $1/x_i$'s, and is a polynomial in $1/N$.
We write:
\beq
\hat\om_n(x_1,\dots,x_n) = \sum_{g=0}^\infty (N/s)^{2-2g-n}\, \hat\om_n^{(g)}(x_1,\dots,x_n)
\eeq
which is an equality between formal series of $s$ (order by order the sum over $g$ is finite).

It was also proved by BIPZ \cite{BIPZ}, that $\hat\om_n^{(g)}$ is the generating function for maps of genus $g$, with $n$ marked faces, and with 1 marked edge on each marked face:
\beq
\hat\om_n^{(g)} = \sum_{{\rm maps,\, genus}\,g} {s^{\#{\rm vertices}}\over \#{\rm Aut}}\,\,\, {s_3^{\# {\rm triangles}}\, s_4^{\# {\rm quadrangles}} \dots \, s_{d+1}^{\# (d+1){\rm -gons}}\over x_1^{l_1+1}\dots x_n^{l_n+1}}
\eeq
where $l_i$ is the length of the $i^{\rm th}$ marked face.

All this is now standard result in combinatorics of maps.

\subsubsection{Spectral invariants and spectral curve}
\label{secspinv1}

More recently, it was proved in \cite{eynloop1mat, EOFg}, that the functions $\hat\om_n^{(g)}$ and $\hat f_g$ are the spectral invariants (in the sense of \cite{EOFg}) of the formal matrix model spectral curve ${\cal E}_{\rm formal\, MM}=(\hat x,\hat y)$ defined parametrically by:
\beq
{\cal E}_{\rm formal\, MM} = 
\left\{
\begin{array}{l}
\hat x(z) = \alpha + \gamma(z+{1\over z}) \cr
\hat y(z) = \sum_{k=1}^d u_k (z^k-z^{-k})
\end{array}
\right.
\eeq
where the coefficients $\alpha, \gamma, u_k$ are entirely determined by the following algebraic constraints:
\beq\label{eqconstraintsgammaalpha}
\hat x(z)-\sum_{j=2}^d s_{j+1}\,  \hat x(z)^j = \sum_{k=1}^d u_k(z^k+z^{-k})
\virg u_1={s\over \gamma} \virg u_0=0,
\eeq
which give an algebraic equation for $\alpha$ and $\gamma$, whose solution we choose  such that $\alpha$ and $\gamma^2$ are formal power series of $s$ starting with:
\beq
\gamma^2 = s + O(s^2)
\virg
\alpha = O(s).
\eeq
(We give the example of quadrangulations below)

\bigskip

Then, from \cite{eynloop1mat, ec1loopF} we have:
\bt (proved in \cite{eynloop1mat, ec1loopF, EOFg})\label{thomngmmformalspinv}:
$\hat\om_n^{(g)}$ and $\hat f_g$ are the spectral invariants (in the sense of \cite{EOFg}) of the formal matrix model spectral curve ${\cal E}_{\rm formal\, MM}=(\hat x,\hat y)$.

The spectral invariants of that curve are defined as follows (see \cite{EOFg}):
\beq
\hat\om_2^{(0)}(\hat x(z_1),\hat x(z_2)) = {1\over (z_1-z_2)^2\,\hat x'(z_1)\hat x'(z_2)} - {1\over (\hat x(z_1)-\hat x(z_2))^2},
\eeq
and with $J=\{\hat x(z_1),\dots,\hat x(z_n)\}$, we have recursively:
\bea
&& \hat \om_{n+1}^{(g)}(J,\hat x(z_{n+1})) \cr
&=& {1\over 2 \hat x'(z_{n+1})}\Res_{z\to \pm 1}\, {z\,\hat x'(z)^2\,dz\over z_{n+1} \, (\hat x(z_{n+1})-\hat x(z))\hat y(z)}\,\, 
\Big( \hat\om_{n+2}^{(g-1)}(J,\hat x(z),\hat x(z)) \cr
&& + \sum_{h=0}^g \sum'_{I\subset J}\hat\om_{1+|I|}^{(h)}(I,\hat x(z))\hat\om_{1+n-|I|}^{(g-h)}(J\setminus I,\hat x(z)) \Big) \cr
\eea
and for $g\geq 2$:
\beq
\hat f_g = {1\over 2-2g} \Res_{z\to \pm 1} \hat\om_1^{(g)}(\hat x(z))\,\hat\Phi(z)\, dz
\eeq
where $\hat\Phi'(z)= \hat y(z)\,\hat x'(z) $, and for $g=1$:
\beq
\hat f_1 = {1\over 24}\,\ln{(\gamma^2\hat y'(1) \hat y'(-1))}
\eeq
and for $g=0$:
\bea
\hat f_0 &=& {1\over 2}\Big( \sum_{j\geq 1} {\gamma^2\over j} (u_{j+1}-u_{j-1})^2 + {2 s \gamma \over j}(-1)^j (u_{2j-1}-u_{2j+1}) \cr
&& \qquad + {3 s^2\over 2} + s^2\, \ln{\left( \gamma^2\over s \right)}   \Big) .\cr
\eea

\et

\bigskip

{\bf Example}, for quadrangulations we have $s_4\neq 0$ and all the other $s_k=0$. That gives:
\beq
{\cal E}_{\rm quadrangulations}=\left\{
\begin{array}{l}
\hat x(z) =  \gamma(z+{1\over z}) \cr
\hat y(z) = {s\over \gamma}(z-{1\over z}) - {s_4\gamma^3}(z^3-z^{-3}) \cr
\cr
\gamma^2 = {1\over 6 s_4}\,(1-\sqrt{1-12 s s_4}),\cr
u_0=u_2=\alpha=0 ,\cr
u_1={s\over \gamma} \virg u_3=-s_4\,\gamma^3.
\end{array}
\right.
\eeq
That gives:
\beq
\hat f_0 = {1\over 2}\Big( {\gamma^2\over 2}(u_3-u_1)^2+{\gamma^2\over 4}(u_1)^2  
- {2s\gamma}(u_1-u_3) + {2s\gamma\over 2 }(u_3) + {3s^2\over 2}+s^2\ln{\gamma^2\over s} \Big),
\eeq
\beq
\hat f_1 = {1\over 12}\,\ln{(2(2s-\gamma^2))},
\eeq
\beq
\hat f_2 =  {178 s^3-465 s^2\gamma^2+420 s\gamma^4-130 \gamma^6\over 6!\,\, s_4^2\,\,(\gamma^2-2s)^5},
\eeq
and so on...
Notice that at $s=s_c={1\over 12 s_4}$, $\hat y$ has a singular branch point $\hat y\sim (\hat x-2\gamma)^{3/2}$, and when $s\to s_c$, $\hat f_g$ diverges as $(s-s_c)^{{5\over 4}(2-2g)}$.
Anticipating on what follows, we see that the exponent $5/4=(p+q)/(p+q-1)$ indeed corresponds to the $(p,q)=(3,2)$ minimal model, of central charge $c=0$, called pure gravity.
In other words, the statistics of large quadrangulations is equivalent to the pure gravity $(3,2)$ conformal minimal model, i.e. Liouville field theory.

\subsubsection{Limits of large maps}

It is well known that the asymptotic large size behavior of a number of objects is related to the singularities  of its generating  series.
Therefore, the number of maps with a large number of vertices (large maps), is governed by the singularities, i.e. the values of $s_c$, such that $\ln Z$ is not analytical at $s=s_c$.
One should consider the singularity $s_c$ closest to the origin, i.e. $|s_c|$ minimal, and see how the $\hat f_g$ and $\hat\om_n^{(g)}$'s diverge at $s\to s_c$.
Thus, the scaling limit $s\to s_c$ of a formal matrix integral near a singularity $s_c$, corresponds to the asymptotics of large discrete surfaces.

\smallskip

For instance, one easily sees from \eq{eqfgnbmaps}, that the expectation value of the number of vertices for maps of genus $g$ is:
\beq\nonumber
<\#{\rm vertices}> = s\,{d\over ds}\, \ln{\hat f_g} 
\eeq
and thus large maps become dominant when $s\to s_c$ a singularity of $\hat f_g$.
Typically, if we have an algebraic singularity of the type $\hat f_g \sim (s_c-s)^{-\alpha_g}\,\, f_g$, the expectation value of the number of vertices is:
\beq\nonumber
<\#{\rm vertices}> \sim {\alpha_g\, s_c\over s_c-s},
\eeq
and we see that $(s_c-s)$, i.e. the distance to critical point, can be thought of as the "mesh size", so that the area (i.e. number of vertices times mesh size) remains finite in the limit.

\smallskip

Another way to say that, is imagine that $\hat f_g$ has an algebraic singularity of type
\beq\nonumber
\hat f_g \sim (s_c-s)^{-\alpha_g}\,\, f_g
\eeq
and notice that 
\beq\nonumber
(1-s/s_c)^{-\alpha_g} = \sum_{v=0}^\infty \pmatrix{-\alpha_g\cr v}\,\, (-s/s_c)^v = \sum_{v=0}^\infty {\Gamma(v+\alpha_g)\over v!\, \Gamma(\alpha_g)}\,\, (s/s_c)^v
\eeq
This means that the (possibly weighted) number of maps of genus $g$ with $v$ vertices, behaves for large $v$ as:
\beq\nonumber
f_g\,\,s_c^{-\alpha_g-v}\,\,{\Gamma(v+\alpha_g)\over v!\, \Gamma(\alpha_g)}
\sim f_g\,\,s_c^{-\alpha_g-v}\,\,{v^{\alpha_g-1}\over \Gamma(\alpha_g)}
\eeq
where we used the large $v$ Stirling asymptotic formula for the $\Gamma$-function.

Similarly, the $s\to s_c$ asymptotics of $\hat \om_n^{(g)}(x_1,\dots,x_n)$ give the enumeration of large maps with $n$ marked faces, and if we also rescale $x_i \to (s_c-s)^{\alpha_{n,g}}\, \td x_i$, we ca also consider large maps with large marked faces.

\smallskip

Therefore, we see that the enumeration of large maps, is asymptotically given by the knowledge of:

\noindent - the exponents $\alpha_{n,g}$ (and $\alpha_g=\alpha_{0,g}$), 

\noindent - the critical point $s_c$ 

\noindent - and the prefactor $f_g$.

\smallskip

$\bullet$ It turns out that the critical point $s_c$ is independent of $n$ and $g$, and it can be easily found from the resolvent $\om_1^{(0)}$, it is not universal, it is strongly model dependent.

$\bullet$ The exponent $\alpha_{n,g}$ turns out to be proportional to $(2-2g-n)$:
\beq
\alpha_{n,g} = (2-2g-n)\, (1-\gamma_{\rm string}/2)
\eeq
where $\gamma_{\rm string}$ is a universal exponent, it depends only on $(p,q)$, and it is one of the exponents computed by the famous KPZ formula. Here, we shall see that it is:
\beq
\gamma_{\rm string} = {-2\over p+q-1}.
\eeq

$\bullet$ The last thing to compute, is the prefactor $f_g$, or $\om_n^{(g)}$.
Here, in this article, we prove in theorem \ref{mainth} the long claim statement that this prefactor is the same as the one computed directly from conformal field theory technics, with the Liouville theory coupled to matter reprensented by a minimal model $(p,q)$ of central charge $c=1-6(p-q)^2/pq$ (notice that the $(3,2)$ model has $c=0$ and thus is called pure Liouville gravity).
In particular, we show that the generating function of the coefficients $f_g$, satisfies the Gelfand-Dikii non linear equation, see \eq{eqgenfgu}.

\subsubsection{Singularities of spectral invariants}
\medskip

One can easily convince oneself that the algebraic equations \eq{eqconstraintsgammaalpha} obeyed by $\alpha$ and $\gamma$, are singular whenever $\hat y'(1)=0$  or $\hat y'(-1)=0$, and then from theorem \ref{thomngmmformalspinv}, one can see that the $\hat f_g$'s and $\hat\om_n^{(g)}$'s diverge whenever $\hat y'(1)=0$ or $\hat y'(-1)=0$, i.e. whenever $\hat y$ doesn't behave as a square root branchpoint.

Let us assume that we fix the parameters $s_k$ and $s=s_c$ such that:
\beq
\hat y(z) \mathop{{\sim}}_{z\to 1} (\hat x(z)-\hat x(1))^{m+{1\over 2}}.
\eeq

This can be obtained for instance if we choose:
\bea
V'(x) = (x-\alpha-2)^m\,\, (T_{m+1}(x-\alpha)-T_m(x-\alpha)) \cr
\cr
s_c=(-\alpha-2)^m\, (T'_{m+1}(-\alpha)-T'_m(-\alpha)) \cr
T_{m+1}(-\alpha)=T_m(-\alpha)
\eea
where $T_m(z+z^{-1})=z^m+z^{-m}$ is the Tchebychev's polynomial of degree $m$.
In that case we have at $s=s_c$:
\beq
\left\{
\begin{array}{l}
\hat x(z) =  z+{1\over z} \cr
\hat y(z) = (z-1)^{2m+1}-({1\over z}-1)^{2m+1}
\end{array}
\right.
\eeq

\medskip

When $s$ is close to $s_c$ but not exactly equal to $s_c$, we have like in \eq{eqdefspcurvelim}:
\beq\label{eqdefspcurvelimbis}
\hat y((s-s_c)^{1\over m+1}\, x) \sim (s-s_c)^{2m+1\over 2m+2}\,\,y(x)\,\, (1+O((s-s_c)^{1\over m+1})\,\,
\virg
y(x) = \sum_{k=0}^m c_k\, x^{k+{1\over 2}}.
\eeq
At $s=s_c$ we have $\hat y\sim \hat x^{m+{1\over 2}}$ and at $s\neq s_c$ we have $\hat y\sim \sqrt{\hat x-\hat x(1)}$.
The value of $m$ and the coefficients $c_k$ depend on which limit of large maps we are interested in. Indeed we may fine-tune the coefficients $s_j$, in order to favor one value of $m$ or another.

Again, theorem 8.1. of \cite{EOFg} implies that:
\bt (theorem 8.1. of \cite{EOFg}):\label{thWnglimMMbis}
If $m>0$:
\beq
\hat f_g \sim (s-s_c)^{(2-2g){2m+3\over 2m+2}}\, f_g,
\eeq
and if $m\geq 0$:
\bea
\hat \om_n^{(g)}((s-s_c)^{1\over m+1}x_1,\dots,(s-s_c)^{1\over m+1}x_n) \cr
 \sim (s-s_c)^{(2-2g-n){2m+3\over 2m+2}-{n\over m+1}}\,\om^{(g)}_n(x_1,\dots,x_n)\,\,\, (1+O((s-s_c)^{{1\over m+1}} )),
\eea
where the $f_g$'s and  $\om_n^{(g)}$'s are the spectral invariants of \cite{EOFg} for the spectral curve $y(x)$ appearing in \eq{eqdefspcurvelimbis}.
\et

\bigskip
It was argued and highly debated, that this limit should be equivalent to the Liouville gravity conformal field theory, coupled to some matter field given by a conformal minimal model $(p,q)$ of central charge $c=1-6{(p-q)^2\over pq}$.
Intuitively, discrete surfaces made of a very large number of small polygons, should give a good approximation of smooth Riemann surfaces...

It was indeed proved that the critical exponents $-\alpha_g=(2-2g){2m+3\over 2m+2}$ are the same (given by KPZ formula \cite{KPZ, Duplantier1, Duplantier2}) as those of the Liouville conformal field theory, but it is only recently that it became possible to compute explicitly partition functions and correlation functions on both sides: on the matrix model side (in particular in the double scaling limit), and in the conformal theory side.

On the Liouville conformal theory side, recent progress was obtained following Zamolodchikov, Belavin, Hosomichi, Ribault, Teschner, ... \cite{Bel1,Bel2,Bel3, Tesch1, Tesch2, Rib2}.

On the matrix model side, recent progress was obtained in \cite{eynloop1mat}, and formalized as a special case of the symplectic invariants of \cite{EOFg}, which allow to compute all correlation functions of all genus.

\medskip
From here, we can repeat all what was said in section \ref{secuniversality}, after theorem \ref{thWngsympinv}.

\bigskip

In this article, we shall show how to apply the spectral invariants method of \cite{EOFg}, for the double scaling limit of matrix models which is expected to coincide with Liouville theory.

We prove that the scaling limits of the matrix model correlation functions, i.e. the generating functions counting discrete surfaces, is indeed the $(p,2)$ reduction of KdV satisfying string equation, i.e. the minimal model $(p,2)$.

\section{Minimal models}

There exists several equivalent definitions of minimal models coupled to gravity.
They correspond to representations of the conformal group in 2 dimensions.
They are classified by two integers $(p,q)$, and their central charge is:
\beq
c = 1 - 6{(p-q)^2\over pq}
\eeq

%If we assume that there is a Liouville graviton coupled to matter in the $(p,q)$ representation, there are a finite number of operators, from which all the observables can be generated.

Some of them have received special names:\\
$\bullet$ $(1,2)=$ Airy, $c=-2$ (related to Tracy-Widom law \cite{TW})\\
%$\bullet$ $(2,1)=$ Pearcey, $c=-2$\\
$\bullet$ $(3,2)=$ pure gravity, $c=0$\\
$\bullet$ $(5,2)=$ Lee-Yang edge singularity, $c=-{22\over 5}$\\
$\bullet$ $(4,3)=$ Ising, $c={1\over 2}$\\
$\bullet$ $(6,5)=$ Potts-3, $c={4\over 5}$\\

Minimal models can also be viewed as finite reductions of the Kadamtsev-Petviashvili (KP) integrable hierarchy of partial differential equations.

\medskip

The case $q=2$ is a little bit simpler to address, and is a reduction of the Korteweg de Vries (KdV) hierarchy.

The KdV hierarchy, and the minimal models $(p,2)$ have generated a huge amount of works, and have been presented in many different (but equivalent) formulations. For instance in terms of a string equation for differential operators, in terms of a Lax pair, in terms of commuting hamiltonians, in terms of Schr\"odinger equation, in terms of Hirota equations, in terms of isomonodromic systems, in terms of Riemann Hilbert problems, in terms of tau functions, in terms of Grasman manifolds, in terms of Yang-Baxter equations, ...etc, see \cite{BBT} for a comprehensive lecture.

All those formulations are equivalent, and let us recall some of the well known features of the $(p,2)$ reduction of KdV (see \cite{BookPDF,BBT}), presented  in a way convenient for our purposes.

\subsection{String equation}

The KdV minimal model $(p,2)$ with $p=2m+1$ can be formulated in terms of two differential operators $P$, $Q$ of respective orders $p$ and $2$, satisfying the string equation:
\beq\label{eqstringPQ}
[P,Q]={1\over N}\,{\rm Id}
\eeq
\beq
Q=d^2-2u(t)
\virg
P= d^p  - p\, u\, d^{p-2} + \dots
\virg d={1\over N}\,{d\over dt}
\eeq
${1\over N}$ is a scaling parameter, which we can send to zero to get the "classical limit".

\bigskip

The general solution of the string equation \eq{eqstringPQ} is of the form:
\beq\label{PsumQj+12+}
P = \sum_{j=0}^m t_j (Q^{j+1/2})_+ 
\virg
t_m=1
\eeq
where $(Q^{j+1/2})_+$ is the unique differential operator of order $2j+1$, such that:
\beq
{\rm order} [ ((Q^{j+1/2})_+)^2 - Q^{2j+1} ] \leq 2j.
\eeq
For example:
\beq
(Q^{1/2})_+=d
\virg
(Q^{3/2})_+=d^3-3ud-{3\dot u\over 2},
\eeq
\beq
(Q^{5/2})_+=d^5-5ud^3-{15\dot u\over 2}\,d^2 - {25\ddot u\over 4}\,d-{45 u^2\over 2}\, d - {15 \over 8}\,\mathop{{u}}^{{\dots}}- {45 u \dot u\over 2}.
\eeq

It is a classical result (see \cite{DGZ}) that it satisfies:
\beq\label{commutQ+QRj}
[(Q^{j-1/2})_+,Q] = {1\over N}\, {d\over dt}(R_j(u(t)))
\eeq
where the right hand side is a function (a differential operator of order $0$), and the coefficients $R_j(u)$ are the Gelfand-Dikii differential polynomials \cite{DGZ}. They can be obtained by the recursion:
\beq\label{defRjGD}
R_0=2
\virg
\dot{R}_{j+1} = -2u \dot{R}_j - \dot{u} R_j + {1\over 4\,N^2} {\mathop{{R}}^{\dots}}_j .
\eeq
The first few of them are:
\beq
\begin{array}{l}
R_0=2 \cr
R_1 = -2u \cr
R_2 = 3\, u^2-{1\over 2\,N^2} \ddot{u} \cr
\displaystyle R_3 =  - 5 u^3 + {5 \over 2 N^2} u \ddot{u} + {5\over 4 N^2} {\dot{u}}^2 - {1\over 8 N^4} {\mathop{{u}}^{.\dots}} \cr
\vdots
\end{array}
\eeq
and in general:
\beq
R_j(u) = {2\,\, (-1)^j\,\, (2j-1)!!\over j!}\,\,u^j +\quad \dots \quad - {2\over (2 N)^{2j-2}}\, u^{(2j-2)}.
\eeq

After substitution of \eq{PsumQj+12+} into the string equation \eq{eqstringPQ},  the property \eq{commutQ+QRj} gives a non-linear differential equation  for the function $u(t)$:
\beq\label{streqq}
\encadremath{
\sum_{j=0}^m t_j R_{j+1}(u) =  t.
}\eeq
Since $R_0=2$, we see that we can identify $t$ with $t=-2 t_{-1}$.

\bigskip

$\bullet$ For instance for Airy $p=1$, this gives:
\beq\label{streqqAiry}
 - 2  u  =  t.
\eeq

$\bullet$ For instance for pure gravity $p=3$, this is the Painlev\'e I equation:
\beq\label{streqqPI}
3\, u^2-{1\over 2 N^2} \ddot{u} - 2 t_0 u  =  t.
\eeq

$\bullet$ For instance for Lee-Yang $p=5$, we have:
\beq\label{streqqLeeYang}
- 5 u^3 + {5\over 2 N^2} u \ddot{u} - {1\over 4 N^2} {\dot{u}}^2 - {1\over 8 N^4} {\mathop{{u}}^{.\dots}} + t_1(3\, u^2-{1\over 2 N^2} \ddot{u}) - 2 t_0 u  =  t.
\eeq

%
%\subsection{Commuting Hamiltonians}

%One has:
%\beq
%(2k+1)\,{\partial u\over \partial t_{k-1}} = {\partial \over \partial t}\, R_{k+1}(u) = [(Q^{k+1/2})_+,Q]
%\eeq
%and thus the Hamiltonians:
%\beq
%H_{k-1} = {\partial\over \partial t_{k-1}} + {1\over k+{1\over 2}}\, (Q^{k+1/2})_+
%\eeq
%commute with $Q$:
%\beq
%[H_k,Q]=0
%\eeq
%In fact they commute together:
%\beq
%[H_i,H_j]=0
%\eeq

%% IMPORTANT
%\beq
%Q_1=d \virg Q_{k+1} = Q_k. Q+{1\over 2} R_k. d -{1\over 4} \dot{R}_k = Q.Q_k + {1\over 2} R_k. d -{5\over 4} \dot{R}_k
%\eeq

%

\subsection{Tau function}

We define the Tau-function $\tau(t,t_0,\dots,t_m)$ and its log, the free energy function $F(t,t_0,\dots,t_m)=\ln\tau(t,t_0,\dots,t_m)$ such that:
\beq
N^{-2}\,\ddot{F}= u .
\eeq
The Tau-function has many other properties, which can be found in textbooks and classical works on the subject \cite{BBT, Kri, krichwitham}, but which are beyond the scope of the present article. 

\subsection{Lax pair}

Consider the following matrices:
\beq
{\cal R}(x,t) = \pmatrix{ 0 & 1 \cr x+2u(t) & 0 },
\eeq
and for any integer $k$:
\beq
{\cal D}_k(x,t) = \pmatrix{A_{k} & B_k \cr C_{k} & -A_{k}},
\eeq
where $A_k(x,t), B_k(x,t), C_k(x,t)$ are polynomials of respective degree $k-1,k,k+1$ in $x$, which are determined by:
\beq
B_k(x,t) = \sum_{j=0}^k x^{k-j}\,\, R_j(u)
\virg
A_{k} = -{1\over 2N} \dot{B}_k
\virg
C_{k} = (x+2u)\,B_k+ {1\over N}\dot{A}_{k}.
\eeq

The recursion relation \eq{defRjGD} implies that $B_k$ satisfies the equation:
\beq
2\dot{u} B_k + 2 (x+2u) \dot{B}_k - {1\over 2N^2} {\mathop{{B}}^{\dots}}_k = -2 \dot R_{k+1}(u)
\eeq
and we see that the matrix ${\cal D}_k(x,t)$ satisfies:
\beq\label{eqLax}
{1\over N}\,{\partial \over \partial t} {\cal D}_k(x,t) + [{\cal D}_k(x,t),{\cal R}(x,t)] = -{2\over N}\dot{R}_{k+1}(u)\, \pmatrix{0 & 0 \cr 1 & 0} ,
\eeq
the right hand side is independent of $x$, and is proportional to ${\partial\over \partial x} {\cal R}(x,t)$.

\subsection{Lax equation}

If we consider $u$ solution of the string equation \eq{streqq}, then, the matrix:
\beq
{\cal D}(x,t) = \sum_{j=0}^m  t_j {\cal D}_j(x,t) \virg t_m=1
\eeq
satisfies the Lax equation:
\beq\label{eqLaxbis}
{1\over N}\,{\partial \over \partial t} {\cal D}(x,t) + [{\cal D}(x,t),{\cal R}(x,t)] = -{2\over N}  {\partial \over \partial x} {\cal R}(x,t)
\eeq
which can also be written as a commutation relation:
\beq\label{eqLaxtotal}
 \left[{2\over N}  {\partial \over \partial x}+{\cal D}(x,t),{\cal R}(x,t)-{1\over N}\,{\partial \over \partial t}\right] = 0
\eeq
This relation means that the operator $ {2\over N}\,{\partial \over \partial x}+{\cal D}(x,t)$ is a Lax operator \cite{BBT}.

\subsection{The differential system}

The Lax equation \eq{eqLaxtotal} is the compatibility condition, which says that the following two differential systems have a common solution $\Psi(x,t)$:
\beq\label{PsicommoneqDR}
{1\over N}\,{d\over dx}\, \Psi(x,t) = -{1\over 2}\,{\cal D}(x,t)\, \Psi(x,t)
\virg
{1\over N}\,{d\over dt}\, \Psi(x,t) = {\cal R}(x,t)\, \Psi(x,t)
\eeq
and $\Psi(x,t)$ is a matrix such that:
\beq\label{defpsiBAmmodel}
\Psi(x,t)= \pmatrix{\psi & \phi \cr \td\psi & \td\phi}
\virg
\det\Psi=1.
\eeq
In particular we have the Schr\"odinger equation for $\psi$:
\beq\label{Scrheodinger}
{1\over N^2}\, \ddot\psi(x,t) = (x+2u(t))\, \psi(x,t)
\eeq
where $t$ can be interpreted as the space variable, and $x$ the energy. $x$ is called the spectral parameter.

\subsection{Correlators}

Consider the Christoffel-Darboux kernel associated to the system ${\cal D}(x)$:
\beq\label{detKCDD}
K(x_1,x_2) = {\psi(x_1)\td\phi(x_2) - \td\psi(x_1)\phi(x_2)\over x_1-x_2}
\eeq

\bd\label{defdetcorrel}

We define the connected correlation functions by the "determinantal formulae":
\beq\label{detKW1}
W_1(x) = \mathop{{\rm lim}}_{x'\to x} K(x,x')-{1\over x-x'} = \psi'(x) \td\phi(x) - \td\psi'(x)\phi(x)
\eeq
and for $n\geq 2$:
\beq\label{detKW}
W_n(x_1,\dots,x_n) = 
 - {\delta_{n,2}\over (x_1-x_2)^2} -(-1)^n\, \sum_{\sigma={\rm cyles}} \prod_{i=1}^n K(x_{\sigma(i)},x_{\sigma(i+1)}) 
\eeq

\ed

For example:
\beq
W_3(x_1,x_2,x_3) = K(x_1,x_2)K(x_2,x_3)K(x_3,x_1)+K(x_1,x_3)K(x_3,x_2)K(x_2,x_1).
\eeq

Although we have not written it explicitly, the kernel $K$ and the correlators $W_n$ depend on $t$.

\medskip

The {\em non-connected} correlation functions are defined by:
\beq
W_{n,\, n.c.}(x_1,\dots,x_n) = \sum_k \sum_{J_1\cup J_2\cup \dots \cup J_k=J} \prod_{i=1}^k W_{|J_i|}(J_i),
\eeq
where $J=\{x_1,\dots,x_n\}$ and the sum runs over all partitions of $J$ into $k$ non-empty disjoint subsets.
In other words, the connected $W_n$'s are the cumulants of the non-connected ones.

For instance:
\beq
W_{2,\,  n.c.}(x_1,x_2) = W_2(x_1,x_2) + W_1(x_1)W_1(x_2),
\eeq
\bea
W_{3,\,  n.c.}(x_1,x_2,x_3) 
&=& W_3(x_1,x_2,x_3) + W_1(x_1)W_2(x_2,x_3) + W_1(x_2)W_2(x_1,x_3) \cr
&& + W_1(x_3)W_2(x_1,x_2) + W_1(x_1)W_1(x_2)W_1(x_3).
\eea
The formula \eq{detKW} is called "determinantal formula", because for the non-connected correlation functions we have:
\beq
W_{n,\, n.c.}(x_1,\dots,x_n) = \det'(K(x_i,x_j)),
\eeq
where $\det'$ means that when we compute the determinant as a sum over permutations of products
$(-1)^\sigma\, \prod_i K(x_i,x_{\sigma(i)})$, then if $\sigma(i)=i$ we replace $K(x_i,x_i)$ by $W_1(x_i)$, and if $\sigma(i)=j$ and $\sigma(j)=i$, we replace $K(x_i,x_j)K(x_j,x_i)$ by $-W_2(x_i,x_j)$, see \cite{bergere}.

For instance $W_{3,\, n.c.}$ is the sum of 6 terms coming from the 6 permutations:
\bea
W_{3,\, n.c.}(x_1,x_2,x_3) 
&=& \det'\pmatrix{K(x_1,x_1) & K(x_1,x_2) & K(x_1,x_3) \cr K(x_2,x_1) & K(x_2,x_2) & K(x_2,x_3) \cr K(x_3,x_1) & K(x_3,x_2) & K(x_3,x_3) \cr}  \cr
&=& W_1(x_1) W_1(x_2) W_1(x_3) + W_1(x_1) W_2(x_2,x_3)+W_1(x_2) W_2(x_1,x_3) \cr
&& +W_1(x_3) W_2(x_1,x_2) + K(x_1,x_2)K(x_2,x_3)K(x_3,x_1)\cr
&& + K(x_1,x_3)K(x_3,x_2)K(x_2,x_1)
\eea

\bigskip

It was proved in \cite{BergEyntopdet}, that the correlators $W_n$ satisfy an infinite set of equations, called loop equations, and equivalent to Virasoro constraints for the $\tau$ function.
The loop equation simply states that the following quantity:
\bt\label{thloopeq1} Loop equations (proved in \cite{BergEyntopdet}):
\bea
&& P_n(x;x_1,\dots,x_n) \cr
&=& W_{n+2,\, n.c.}(x,x,x_1,\dots,x_n) \cr
&& + \sum_{j=1}^n {\partial \over \partial x_j}\, {W_n(x,x_1,\dots,x_{j-1},x_{j+1},\dots,x_n)-W_n(x_1,\dots,x_n)\over x-x_j}  \cr
\eea
is a polynomial of the variable $x$.
\et

For example, one can easily check that:
\beq
P_0(x)= W_2(x,x)+W_1(x)^2 = - \det {\cal D}(x,t) .
\eeq
Notice that
\beq
{1\over N}\,{\partial\over \partial t}\, \det {\cal D}(x,t) = 2 B(x,t) = 2\sum_{j=0}^m t_j B_j(x,t).
\eeq

\subsection{Example: Airy kernel}

Let us write the $(1,2)$ model, i.e. $m=0$.
We have:
\beq
P= d \virg Q=d^2-2u
\eeq
the string equation is:
\beq
 [P,Q] = -{2\over N}\,\, \dot{u}= {1\over N}
\eeq
i.e.
\beq
u(t) = -{t\over 2} = t_{-1}
\eeq
The Lax pair is:
\beq
{\cal D}_0(x,t) = \pmatrix{ 0 & 2 \cr 2x+4u & 0}
\virg
R(x,t) = \pmatrix{ 0 & 1 \cr x+2u & 0}
\eeq
The differential system is:
\beq
{1\over N}\,{d\over dx}\, \Psi(x,t) = -\pmatrix{ 0 & 1 \cr x-t & 0} \, \Psi(x,t)
\eeq
i.e.
\beq
\psi'' = N^2 (x-t) \psi
\eeq
whose solution is the Airy function \cite{Abramowitz}:
\beq
\psi(x,t) = Ai(N^{2\over 3}(x-t))
\virg
\td\psi(x,t) = -Ai'(N^{2\over 3}(x-t))
\eeq
and the other independent solution is the "Bairy" function \cite{Abramowitz}:
\beq
\phi(x,t) = Bi(N^{2\over 3}(x-t))
\virg
\td\phi(x,t) = -Bi'(N^{2\over 3}(x-t))
\eeq
and thus the kernel is the famous Airy kernel \cite{TWAiry}:
\beq
K_{\rm Airy}(t+N^{-2/3} x_1,t+N^{-2/3} x_2) = {Ai'(x_1)Bi(x_2)-Ai(x_1)Bi'(x_2)\over x_1-x_2}
\eeq
The Airy kernel plays a very important role in many problems, in particular in the universal laws of extreme values, related to the Tracy-Widom law \cite{TW}.

The $\tau$ function is simply:
\beq
\tau = \ee{-{N^2 t^3\over 12}}.
\eeq

For the Airy system, the polynomial of theorem \ref{thloopeq1} is simply:
\beq
P_n(x) = 4(x+2u)\,\delta_{n,0}.
\eeq

\subsection{Classical limit}

The classical limit is the large $N$ limit, or equivalently, it is also the large $t$ limit.

Intuitively, in the classical limit, $P$ and $Q$ commute, and they can be represented without differential operators. In this limit $d\to z$ can be represented as a number, and operators $Q=d^2-2u$ and $P$ are replaced by functions of $z$ and $t$. Therefore, in analogy with $Q=d^2-2u(t)$, and $P=d^p+\dots$, let us define two functions $x(z,t)$ and $y(z,t)$:
\beq
 x(z,t)=z^2-2u_0(t)
\virg
 y(z,t)=z^p+\dots\,\,.
\eeq
In the classical limit, we replace the string equation $[P,Q]=N^{-1}$  with a Poisson bracket:
\beq\label{xyPoisson}
\{y,x\}=1 = {\partial y\over \partial z}\,{\partial x\over \partial t}-{\partial y\over \partial t}\,{\partial x\over \partial z}
\eeq
whose general solution is:
\beq
x(z,t)=z^2-2u_0(t)
\virg
y(z,t)=\sum_{j=0}^m t_j\, \left( z^{2j+1}\, (1-{2u_0(t)\over z^2})^{j+1/2}\right)_+,
\eeq
where $()_+$ means the positive part in the large $z$ Laurent series expansion. Explicitly we get:
\beq\label{eqdefyzt}
y(z,t) = \sum_{j=0}^m \sum_{l=0}^j t_j z^{2j+1-2l}\, (-u_0/2)^l\,\,  {(2j+1)!\over j!}\,{(j-l)!\over l!\,(2j+1-2l)!}.
\eeq
The string equation $\{y,x\}=1$ reduces to:
\beq
\dot u_0 \,y'(0) = {-1\over 2},
\eeq
i.e.
\beq
\sum_{j=0}^m  t_j \, \dot u_0\,(-u_0/2)^j \,\,  {(2j+1)!\over (j!)^2} = -{1\over 2}
\eeq
which can be integrated with respect to $t$ and gives a polynomial equation for $u_0(t)$:
\beq\label{polequ0}
{\cal P}(u_0) =  \sum_{j=0}^m  t_j \, (-u_0/2)^{j+1}\,\,\, {(2j+1)!\over j!\, (j+1)!} = {t\over 4}
\eeq
which is clearly the classical limit of \eq{streqq}.
In other words, the non-linear differential equation \eq{streqq} for $u(t)$, becomes an algebraic equation for $u_0(t)$.

\medskip

For example, for pure gravity $m=1$ we have the classical limit of \eq{streqqPI}:
\beq\label{streqclPI}
4 {\cal P}(u_0)= 3\, u_0^2 - 2 t_0\, u_0  =  t.
\eeq

\subsection{Topological expansion}

We now have the polynomial equation \eq{polequ0}:
\beq
{\cal P}(u_0)=t/4
\eeq
which implies:
\beq
\dot{u}_0 = {1\over 4{\cal P}'(u_0)}
\virg
\ddot u_0 = {- {\cal P}''(u_0)\over 16\, ({\cal P}'(u_0))^3}
\virg \dots
\eeq
and in general, any derivative of $u_0$ with respect to $t$ can be written as a rational function of $u_0$.

Since $u_0(t)$ satisfies the string equation \eq{streqq} at $N=\infty$, the full solution $u(t)$ to the string equation \eq{streqq}, can be expanded as an $N^{-2}$ power series:
\beq
u(t) = u_0 + \sum_k N^{-2k} \, u_k(t)
\eeq
where all coefficients $u_k$ are rational functions of $u_0$ (their denominator is a power of ${\cal P}'(u_0)$).

For example for pure gravity $m=1$, the Painlev\'e equation \eq{streqqPI} implies that to the first few orders we have:
\beq
u(t) = u_0 -{3\over N^2}\,(6u_0-2t_0)^{-4} +  O(N^{-4}).
\eeq
And the Free energy $F(t)$ such that $u={1\over N^2}\,\ddot{F}$, also has a $1/N^2$ expansion:
\beq
\ln\tau = F = \sum_{g=0}^\infty N^{2-2g} F_g(u_0) \virg \ddot F_g = u_g.
\eeq

Also, since the coefficients of the differential system ${\cal D}(x,t)$ depend on $u(t)$, the matrix ${\cal D}(x,t)$ has a $1/N^2$ expansion:
\beq
{\cal D}(x,t) = \sum_g N^{-2g} {\cal D}^{(g)} (x,t)
\eeq
To leading order we have:
\beq\label{leadingD}
{\cal D}^{(0)} (x,t) = \pmatrix{0 \qquad & \quad \ovl{B}(x,u_0) \cr  (x+2u_0)\,\ovl{B}(x,u_0) & 0}
\eeq
\beq
\ovl{B}(x,u_0) = 2\,\sum_{j=0}^m\sum_{k=0}^j t_j x^{j-k}\,u_0^k \,\,{(-1)^k\,\, (2k-1)!!\over k!}
\eeq
Notice that:
\beq
z\, \ovl{B}(z^2-2 u_0,u_0) = y(z,t).
\eeq

The classical spectral curve is given by the eigenvalues of ${\cal D}^{(0)}(x,t))$, i.e. the values of $y$ such that $\det{(y-{\cal D}^{(0)}(x,t))}=0$, i.e., if we parametrize $x$ as $x=z^2-2u_0$, we have:
\beq
y =  \pm\, y(z,t)
\eeq
where $y(z,t)$ is the function defined in \eq{eqdefyzt}.
This explains why we call the function $y(z,t)$ the {\em classical spectral curve}.

\medskip
Written in a parametric form where $u_0=u_0(t)$, the classical spectral curve is thus:
\beq\label{spcurvep2}
\encadremath{
{\cal E}_{(2m+1,2)} = \left\{
\begin{array}{l}
x(z) = z^2-2u_0 \cr
y(z) = \sum_j \sum_l t_j z^{2j+1-2l}\, (-u_0/2)^l\,\, \, {(2j+1)!\over j!}\,{(j-l)!\over l!\,(2j+1-2l)!}
\end{array}
\right.
}\eeq
It is important to notice that it is a genus $0$ hyperelliptical curve, which is equivalent to saying that it can be parametrized by a complex variable $z$ (higher genus would be parametrized by a variable $z$ living on a Riemann surface), and which is equivalent to saying that the polynomial $y^2$, written as a polynomial in $x$, has only one simple zero, located at $x=-2u_0$, all the other zeroes are double zeroes.

\subsection{BKW expansion}
\label{secBKW}

Similarly, we can look for a BKW asymptotic solution of the solutions $\psi(x,t)$ of the differential system.
It takes the form:
\beq
\psi(x,t) \sim \,\, {\ee{N\int^x_{-2u_0} ydx}\over \sqrt{2}\, (-x-2 u_0)^{1\over 4}}\,\, \left(1+\sum_k N^{-k} \psi_k(x,u_0) \right)
\eeq
\beq
\td\psi(x,t) \sim \,\, {\ee{N\int^x_{-2u_0} ydx}\,\, (x+2 u_0)^{1\over 4}}\,\, \left(1+\sum_k N^{-k} \td\psi_k(x,u_0) \right)
\eeq
and we recall that $z=(x+2 u_0)^{1\over 2}$.
the BKW expansion of the other solutions $\phi$ and $\td\phi$, are obtained by changing the sign of the square root $z\to -z$.

We have the following Lemma:

\bl
Each $\psi_k(x,u_0)$ and $\td\psi_k(x,u_0)$ is a polynomial of  $1/z$.

\el

\proof{
The proof uses the Schr\"odinger equation \eq{Scrheodinger}:
\beq
{1\over N^2} \ddot\psi(x,t) = (x+2u(t))\,\psi(x,t).
\eeq
Let us write:
\beq
\psi(x,t) = \sqrt{f(x,t)}\, \ee{\int^t {dt'\over f(x,t')}}.
\eeq
The Schr\"odinger equation implies that:
\beq
N^2 (x+2u(t))\,f^2(x,t) = {1\over 2} f(x,t) \ddot f(x,t) - {1\over 4} \dot f(x,t)^2 +1,
\eeq
and after differentiating once more with respect to $t$, we obtain a third order linear equation for $f$:
\beq
(x+2u(t))\,\dot f(x,t) + \dot u(t)\,f(x,t) = {1\over 2\,N^2}\,\, \mathop{f}^{\dots}(x,t).
\eeq
To leading order we have $u(t)=u_0(t)$, and recall that $u(t)$ has a $1/N^2$ expansion, therefore, one easily sees that:
\beq
f(x,t) = {-1\over N\,\sqrt{x+2u_0(t)}}\,\, \Big(1 + \sum_k N^{-2k} \, f_k(x,t)\Big),
\eeq
and by an easy recursion, we see that each $f_k(x,t)$ is a polynomial in $1/z$ with $z=\sqrt{x+2u_0(t)}$. 

Then, notice that the Poisson equation \eq{xyPoisson} implies:
\beq\label{dydtclassical}
\left. {\partial\,y \over \partial t} \right|_{x} = - {1\over x'(z)} = - {1\over 2z} = - {1\over 2\sqrt{x+2u_0}}
\eeq
And therefore:
\beq
\left. {\partial\,\int^x ydx \over \partial t} \right|_{x} = - z .
\eeq
This implies that:
\beq
\int^t {1\over f(x,t)} = N \int^x ydx + \sum_{k\geq 1} N^{1-2k}\, g_k(x,t)
\eeq
and where all coefficients $g_k(x,t)$ are polynomials of $1/z$.

Since $\psi(x,t) = \sqrt{f(x,t)}\, \ee{\int^t {dt'\over f(x,t')}}$, we find that $\psi(x,t)$ is of the form:
\beq
\psi(x,t) \sim \,\, {\ee{N\int^x ydx}\over (x+2 u_0)^{1\over 4}}\,\, \left(1+\sum_k N^{-k} \psi_k(x,u_0) \right)
\eeq
where each $\psi_k(x,t)$ is a polynomial in $1/z$.

The proof for $\td\psi(x,t)$ works in a similar manner.
}

\medskip

This lemma implies that the kernel also have a $1/N$ expansion:
\beq
K(z_1,z_2) = {\ee{N\int^{z_1}_{z_2} ydx}\over 2\,\sqrt{z_1 z_2}\, (z_1-z_2)}\,\Big(1+ \sum_k N^{-k} K_k(z_1,z_2) \Big),
\eeq
where each $K_k(z_1,z_2)$ is a polynomial in $1/z_1$ and in $1/z_2$.

\medskip

This implies that the correlators also have a $1/N$ expansion:
\bl\label{lemmatopexpz}
\beq
W_n(x_1,\dots, x_n) = \sum_g N^{2-2g-n}\, W_n^{(g)}(x_1,\dots,x_n)
\eeq
where each $W_n^{(g)}$ is a rational function of the $z_i=\sqrt{x_i+2u_0}$, with poles only at $z_i=0$, except $W_2^{(0)}$ and $W_1^{(0)}$ which are:
\beq
W_1^{(0)} = y(z,t)
\eeq
\beq
W_2^{(0)} = {1\over 4 z_1 z_2}\,\,{1\over (z_1-z_2)^2} \,\, - {1\over (z_1^2-z_2^2)^2} = {1\over 4 z_1 z_2 (z_1+z_2)^2}.
\eeq
\el

The important point, is that each $W_n^{(g)}$ has no other pole than $z_i=0$, in particular, has no pole at the other zeroes of $y(z,t)$.

\proof{
Notice that in the products $\prod_i K(z_{\sigma(i),\sigma(i+1)})$, all the exponentials cancel, and the result is, order by order in $N^{-k}$, a rational fraction of the $z_i$'s having poles at $z_i=0$, or at $z_i=z_j$.
Except for $W_1^{(0)}$ and $W_2^{(0)}$, the poles at $z_i=z_j$ are simple poles, and it is easy to see that in the sum over permutations, the residues cancel, therefore, each $W_n^{(g)}$ is a rational function of the $z_i$'s having poles only at $z_i=0$.
The cases of $W_2$ and $W_1$ need to be treated separately, and are easy.

\smallskip

The fact that $W_n$ has a $1/N^2$ expansion instead of $1/N$ comes from a simple symmetry argument.
In the expression of $W_n$, changing $\psi\to \phi$ and $\td\psi\to \td\phi$, can also be obtained as changing the order of the $x_i$'s, and since we take a symmetric sum, only the terms which are invariant under the exchange $\psi\to \phi$ and $\td\psi\to \td\phi$ contribute to $W_n$.
Exchanging the two solutions $\psi\to \phi$ and $\td\psi\to \td\phi$, is also equivalent to changing $N\to -N$, and therefore $W_n$ has a given parity in $N$.
}

\subsection{Symplectic invariants}

It was found in \cite{BergEyntopdet}, that the correlators obtained from the determinantal formulae \eq{detKW1}, \eq{detKW} of a Christoffel-Darboux kernel $K$ of type \eq{detKCDD}, do satisfy loop equations, i.e. for any $n$ and $g$, and $J=\{x_1,\dots,x_n\}$, the following quantity:
\bea\label{loopeq}
P_n^{(g)}(x;J) 
&=& \sum_{h=0}^g\sum_{I\subset J} W_{1+|I|}^{(h)}(x,I)W_{1+n-|I|}^{(g-h)}(x,J/I) \cr
&& + \sum_{j=1}^n {\partial \over \partial x_j}\,{W_{n}^{(g)}(x,J/\{x_j\})-W_{n}^{(g)}(x_j,J/\{x_j\})\over x-x_j} 
\eea
is a polynomial in $x$. 
This property, as was proved in \cite{BergEyntopdet}, is a direct consequence of \eq{detKCDD} and \eq{detKW1}, \eq{detKW}.

\smallskip
Moreover we know from section \ref{secBKW}, that $W_n^{(g)}(x(z_1),\dots,x(z_n))$
the following differential form:
\beq\label{omngWngpoles}
{\cal W}_n^{(g)}(z_1,\dots,z_n) = W_n^{(g)}(x(z_1),\dots,x(z_n))\, x'(z_1)\dots x'(z_n) + {\delta_{n,2}\delta_{g,0}x'(z_1)x'(z_2)\over (x(z_1)-x(z_2))^2}
\eeq
is a symmetric rational function of all its variables, and if $2g+n-2>0$, due to lemma \ref{lemmatopexpz}, it has poles only at $z_i=0$,
and
\beq
{\cal W}_2^{(0)}(z_1,z_2) = {1\over (z_1-z_2)^2}
\eeq

\medskip

It was found in \cite{eynloop1mat}, that the unique solution of loop equations \eq{loopeq} which has a topological expansion for which the ${\cal W}_n^{(g)}$'s have the poles given by lemma \ref{lemmatopexpz}, can be obtained by the following recursion relation:
\bt
\bea\label{defrecursivspinvomng}
{\cal W}_{n+1}^{(g)}(z_1,\dots,z_n,z_{n+1}) 
&=& {-1\over 4}\,\Res_{z\to 0}\, {dz\over (z_{n+1}^2-z^2)\,\, y(z)}\, \Big[ {\cal W}_{n+2}^{(g-1)}(z,-z,J) \cr
&& + \sum_{h=0}^g\sum'_{I\subset J} {\cal W}_{1+|I|}^{(h)}(z,I){\cal W}_{1+n-|I|}^{(g-h)}(-z,J/I) \Big]
\eea
where $J=\{z_1,\dots,z_n\}$, and $\sum_h\sum'_I$, means that we exclude the terms $(h,I)=(0,\emptyset)$ and $(h,I)=(g,J)$.
\et

\proof{The proof proceeds exactly like in \cite{eynloop1mat}.
Write the Cauchy residue formula:
\beq
{\cal W}_{n+1}^{(g)}(z_1,\dots,z_n,z_{n+1}) = \Res_{z\to z_{n+1}} {dz\over z-z_{n+1}}\,\, {\cal W}_{n+1}^{(g)}(z_1,\dots,z_n,z) 
\eeq
and move the integration contour, to enclose all the other poles, i.e. only $z=0$, and thus:
\bea
{\cal W}_{n+1}^{(g)}(z_1,\dots,z_n,z_{n+1}) 
&=& \Res_{z\to 0} {dz\over z_{n+1}-z}\,\, {\cal W}_{n+1}^{(g)}(z_1,\dots,z_n,z) \cr
&=& \Res_{z\to 0} {x'(z)\, dz\over z_{n+1}-z}\,\, W_{n+1}^{(g)}(x(z_1),\dots,x(z_n),x(z)) \cr
\eea
Then, insert in the right hand side \eq{loopeq}:
\bea
-2W_1^{(0)}(x){W}_{n+1}^{(g)}(x_1,\dots,x_n,x) 
&=& \sum_{h=0}^g\sum'_{I\subset J} { W}_{1+|I|}^{(h)}(x,I){ W}_{1+n-|I|}^{(g-h)}(x,J/I) \cr
&& + \sum_{j=1}^n {\partial \over \partial x_j}\,{{ W}_{n}^{(g)}(x,J/\{x_j\})-{ W}_{n}^{(g)}(x_j,J/\{x_j\})\over x-x_j} \cr
&& - P_n^{(g)}(x;x_1,\dots,x_n)
\eea

and notice that the polynomial $P_n^{(g)}$ has no pole and doesn't contribute to the residue.
All what remains is \eq{defrecursivspinvomng}.
}

\medskip
The recursion relation \eq{defrecursivspinvomng} is precisely the definition of the symplectic invariant's correlators defined in \cite{EOFg}.
In \cite{EOFg}, it is explained how to associate an infinite family of ${\cal W}_n^{(g)}$'s, to any spectral curve defined by a pair of functions $(x(z),y(z))$.

\medskip
Examples:  

$\bullet$ \eq{defrecursivspinvomng} gives:
\beq
{\cal W}_3^{(0)}(z_1,z_2,z_3) = {1\over 2\,y'(0)}\,\, {1\over z_1^2\, z_2^2\, z_3^2}
\eeq

$\bullet$ \eq{defrecursivspinvomng} gives:
\bea
{\cal W}_1^{(1)}(z_1) 
&=&  {-1\over 4}\,\Res_{z\to 0}\, {dz\over (z_{1}^2-z^2)\,\, y(z)}\, {1\over 4 z^2} \cr
&=&  {-1\over 32}\,\, {d^2\over dz^2}\,\left({z\over (z_{1}^2-z^2)\,\, y(z)}\right)_{z=0} \cr
&=& {y'''(0) \over 48\, y'(0)^2}\,{1\over z_1^2} - {1\over 16\, \, y'(0)}\, {1\over z_1^4} 
\eea

\subsection{Double scaling limit and $(p,2)$ kernel}

By comparison with theorem \ref{thWnglimMM}, we conclude that:
\bt\label{mainth}
the $s\to s_c$ double scaling limit of (possibly formal) matrix integrals correlation functions $W_n^{(g)}$ are the determinantal formula correlation functions of the $(p,2)$ kernel:
\beq
\encadremath{
\om_n^{(g)}(x_1,\dots,x_n) = W_n^{(g)}(x_1,\dots,x_n).
}\eeq
where
\beq
\om_n^{(g)}(x_1,\dots,x_n) = \mathop{{\rm lim}}_{s\to s_c} 
(s-s_c)^{(2g+n-2){2m+3\over 2m+2}+{n\over m+1}} \hat \om_n^{(g)}((s-s_c)^{1\over m+1}x_1,\dots,(s-s_c)^{1\over m+1}x_n) 
\eeq
is the double scaling limit of matrix integrals correlators, and 
$W_n^{(g)}(x_1,\dots,x_n)$ is the $g^{\rm th}$ term in the BKW expansion of the determinantal correlator of the $(2m+1,2)$ minimal model (def \ref{defdetcorrel}).

And similarly if $m>0$:
\beq
\mathop{{\rm lim}}_{s\to s_c} (s-s_c)^{(2g-2){2m+3\over 2m+2}} \hat f_g = f_g = F_g
\eeq
where $\hat f_g$ are the free energies of the matrix model and $F_g$ are the free energies of the $(2m+1,2)$ minimal model:
\beq
\ln \tau = \sum_g N^{2-2g} F_g.
\eeq
The double scaling limit is:
\beq
N\to\infty\virg s\to s_c \virg (s-s_c)N^{2m+2\over 2m+3} = O(1) .
\eeq

\et

Therefore, we have proved that, as announced, the double scaling limit of matrix models, is given by the Liouville minimal models $(2m+1,2)$ coupled to gravity.

\subsection{Parametrics of orthogonal polynomials and Baker-Akhiezer functions}

Many approaches of matrix models use some orthogonal polynomials (see Mehta \cite{Mehtabook}):
\beq
p_n(x) = x^n+\dots
\virg
\int p_n(x)\,p_m(x)\,\ee{-{N\over s}V(x)}\,dx = h_n\, \delta_{n,m},
\eeq
which we prefer to make orthonormal:
\beq
\psi_n(x) = {\ee{-{N\over 2s}V(x)}\over \sqrt{h_n}}\,\, p_n(x),
\eeq
as well as their Hilbert transforms:
\beq
\phi_n(x) = \ee{{N\over 2s}V(x)}\, \int {\psi_n(x')\,\,\ee{-{N\over 2s}V(x')} \over x-x'}\,\,dx'.
\eeq 
The matrix
\beq
\Psi_n(x) = \pmatrix{\psi_n(x) & \phi_n(x) \cr \psi_{n-1}(x) & \phi_{n-1}(x)}
\eeq
satisfies a $2\times 2$ differential system ${\cal D}_n(x)$ with polynomial coefficients of degree at most $\deg V'$:
\beq
{1\over N} {d\over dx} \Psi_n(x) = {\cal D}_n(x)\, \Psi_n(x).
\eeq
It is easy to see that $\tr {\cal D}_n(x)=0$ and $\det \Psi_n(x)= \sqrt{h_{n-1}/h_n}$ is constant.

Since $\phi_n(x)$ is discontinuous across the integration contour of $dx'$, the matrix $\Psi_n(x)$ is also discontinuous, and has jumps of the form:
\beq
\Psi_{n\, +}(x) = \Psi_{n\, -}(x)\,\,\pmatrix{1 & 2i\pi \cr 0 & 1}.
\eeq
Therefore $\Psi_n(x)$ satisfies an isomonodromic Riemann-Hilbert problem (the jump matrix, called the monodromy, is independent of $n$ and of $V(x)$). 

A general method was invented \cite{DKMVZ} to find large $N$ asymptotic solution of isomonodromic  Riemann-Hilbert problems.
In the case where we approach a $(2m+1,2)$ singularity, the method of Deift\& co \cite{DKMVZ} requires to have an ansatz for a parametrix asymptotics for $\Psi_n(x)$ in the vicinity of the singularity.

We claim that the correct parametrix for the $(2m+1,2)$ singularity, is the matrix of Baker-Akhiezer functions of \eq{defpsiBAmmodel} for the $(2m+1,2)$ minimal model:
\beq
\Psi_n((s-s_c)^{1\over m+1} x) \sim \Psi(x) \,\, (1+O((s-s_c)^{1\over m+1})).
\eeq

This should be checked by the steepest descent Riemann-Hilbert method of \cite{DKMVZ}.

\section{Kontsevich's integral}

In this section, we also propose a combinatorical interpretation of the coefficients of the $f_g$'s and $\om_n^{(g)}$'s, based on the comparison with Kontsevich integral.

\bigskip

The spectral curve \eq{spcurvep2} of the $(2m+1,2)$ minimal model
\beq
{\cal E}_{(2m+1,2)} = \left\{
\begin{array}{l}
x(z) = z^2-2u_0 \cr
y(z) = \sum_j \sum_l t_j z^{2j+1-2l}\, (-u_0/2)^l\,\, \, {(2j+1)!\over j!}\,{(j-l)!\over l!\,(2j+1-2l)!}
\end{array}
\right.
\eeq
is of the same form as the Kontsevich integral's spectral curve (see in \cite{eynkappa}, or see below), and thus it has the same correlators and spectral invariants $F_g$ as those of the Kontsevich integral.
Therefore, the correlators and $F_g$'s of the minimal model $(2m+1,2)$ can be written as integrals of tautological classes on the moduli spaces of Riemann surfaces.
This can be viewed as a proof that the double scaling limits of matrix models, i.e. the limit of large maps generating function, indeed coincides with topological gravity, as claimed by Witten \cite{witten, DReview} and then proved by Kontsevich \cite{Konts}.

\medskip

The Kontsevich integral \cite{Konts}:
\beq
Z_K(\L) = \int dM \ee{-N\Tr {M^3\over 3} - M\L^2} = \ee{\sum_g N^{2-2g} F_g}
\virg
\tau_k = {1\over N}\Tr \L^{-k}
\eeq
(to simplify we assume $\tau_1=0$ here)
is the generating function for intersection numbers of cotangent line bundles at marked points of Riemann surfaces of genus $g$:
\beq
F_g = W_0^{(g)} = \sum_{\sum_i d_i=3g-3}\, \prod_i {\hat\tau_i^{d_i} \over d_i!}\,\, \left<\prod_i \psi_i^{d_i} \right>_{\overline{{\cal M}}_g}
\virg \psi_i=c_1({\cal L}_i)
\eeq
where $\psi_i$ is the Chern class of the cotangent line bundle at point $i$, and
where
\beq
\hat\tau_i = (2i-1)!!\,\, \tau_{2i+1}
\eeq
and more generally, correlation functions of the Kontsevich integral give access \cite{eynkappa} to integrals of Mumford $\kappa$ characteristic classes  \cite{mumford}:
 \bea
  W_{n}^{(g)}(z_1,\dots,z_n)  
&=& 2^{- d_{g,n}}(\tau_3-2)^{2-2g-n}\!\!\!\!  \sum_{d_0+d_1+\dots+d_n=d_{g,n}}\, 
 \sum_{k=1}^{d_0} {1\over k!}\,\sum_{b_1+\dots+b_k =d_0, b_i>0}  \cr
 && \qquad \qquad \prod_{i=1}^n {2d_i+1!\over d_i!}\, {dz_i\over z_i^{2d_i+2}}\,\, \prod_{l=1}^k \td\tau_{b_l} <\prod_{l=1}^k \kappa_{b_l} \prod_{i=1}^n  \psi_i^{d_i}>_{g,n}  \cr
\eea
The class $\kappa_0$ is the Euler class, and $2\pi \kappa_1$ is the curvature form of the Weil-Petersson symplectic metrics.
The dual times $\td\tau_k$ are closely related to the $\tau_k$'s, see the relation in \cite{eynkappa} or \eq{eqtdtauktauk} below.

It was shown in \cite{EOFg,eynkappa}, that Kontsevich's integral's $W_0^{(g)}=F_g$'s and correlators $W_n^{(g)}$ can be computed as the symplectic invariants $F_g = F_g({\cal E}_K)$ of a spectral curve:
\beq\label{spcurveKonts}
\encadremath{
{\cal E}_{{\rm K}} = \left\{
\begin{array}{l}
x(z) = z^2 \cr
y(z) = z-{1\over 2} \sum_j \tau_{j+2} z^j
\end{array}
\right.
}\eeq

\bigskip

We see that the minimal model ${\cal E}_{(p,2)}$ spectral curve \eq{spcurvep2} can be identified with Kontsevich integral's spectral curve ${\cal E}_{\rm K}$, under the identification of times:
\beq
\delta_{k,0}-{1\over 2}\,\tau_{2k+3} =  {k!\over (2k+1)!}\, \sum_l t_{l+k} \, (-u_0/2)^l\,\,\, {(2l+2k+1)!\over l!\,(l+k)!}
\eeq

In particular
\beq
1-{1\over 2}\,\tau_3 = y'(0) = {1\over -2\,\dot u_0}
\eeq

The dual times $\td\tau_k$ are given by their generating function $\td{g}(r) = \sum_k \td\tau_k r^k$ and $\td{g}(r) =-\ln{(1-g(r))}$ with:
\beq\label{eqtdtauktauk}
1-g(r) = -2 \dot u_0\,\, \sum_{k\geq 0} r^k\, \sum_l t_{l+k} \, (-u_0/2)^l\,\,\, {(2l+2k+1)!\over l!\,(l+k)!} = \ee{-\td{g}(r)}
\eeq
\beq
\td{g}(r) = \sum_k \td\tau_k r^k
\eeq
I.e.
\bea
1-g(r) 
&=& -2\,\dot u_0\,\, \sum_j \sum_{l=0}^j t_{j}\,r^{j-l} \, (-u_0/2)^l\,\, \,\, {(2j+1)!\over l!\, j!} \cr
&=& -2\,\dot u_0\,\, \sum_j  {(2j+1)!\over  j!}\, t_{j} \, r^j\, \sum_{l=0}^j \, {1\over l!}\, (-u_0/2r)^l \cr
&=& -2\,\dot u_0\,\, \sum_j  {(2j+1)!\over  j!}\, t_{j} \, \left( r^j\, \ee{-u_0/2r}\right)_+ \cr
\eea

\section{Derivatives}

The general method to compute derivatives of $F_g$ and $W_n^{(g)}$'s with respect to any parameter entering the spectral curve is explained in \cite{EOFg}.

Here, our spectral curve ${\cal E}_{(p,2)}$ depends on the parameters $t_j$'s and $t=-2 t_{-1}$.
\cite{EOFg} says that we first have to study the variation of  $y(z) x'(z)$ under variation of any such parameter, and we write it:
\beq
{\partial y(z)\over \partial t_j}\, x'(z) - {\partial x(z)\over \partial t_j}\, y'(z) = \L_j'(z)
\eeq
Here, we find for $j\geq 0$:
\beq
\L_j(z) =  -2 {(2j+1)!\over j!}\,u_0\,\,\sum_l z^{2j+1-2l}\, (-u_0/2)^l\,\, \, {(j-l)!  \over (l+1)!\,(2j+1-2l)!}
\eeq
and for $j=-1$:
\beq
\L_{-1}(z) = -2\dot u_0\, y(z) = -{\dot u_0\over z}\,\sum_{j=0}^m t_j \L_j'(z)
\eeq

The theorem 5.1 of \cite{EOFg}, then shows that those functions are such that:
\beq
{\partial W_n^{(g)}(z_1,\dots,z_n) \over \partial t_j} = \Res_\infty W_{n+1}^{(g)}(z_1,\dots,z_n,z_{z+1})\, \L_j(z_{n+1})\,x'(z_{n+1})
\eeq
and in particular for $n=0$:
\beq
{\partial F_g \over \partial t_j} = \Res_\infty W_{1}^{(g)}(z)\, \L_j(z)\,x'(z) = \Res_\infty {\cal W}_{1}^{(g)}(z)\, \L_j(z)
\eeq
and:
\beq
{\partial^k F_g \over \partial t_{j_1} \dots \partial t_{j_k}} = \Res_\infty \dots \Res_\infty {\cal W}_{k}^{(g)}(z_1,\dots,z_k)\, \L_{j_1}(z_1)\dots \L_{j_k}(z_k)
\eeq

\subsubsection{Example}

The recursion relation \eq{defrecursivspinvomng} gives:
\beq
{\cal W}_3^{(0)}(z_1,z_2,z_3)= W_3^{(0)}(z_1,z_2,z_3)\,x'(z_1)x'(z_2)x'(z_3) = {1\over 2y'(0)}\, {1\over z_1^2\,z_2^2\,z_3^2}
\eeq
This implies:
\bea
{\partial^3 F_0\over \partial t_{j_1}\partial t_{j_2}\partial t_{j_3}}
&=& \Res_\infty\Res_\infty\Res_\infty {\cal W}_3^{(0)}(z_1,z_2,z_3)\, \L_{j_1}(z_1)\,\L_{j_2}(z_2)\,\L_{j_3}(z_3) \cr
&=& {-1\over 2y'(0)}\,\,  \L_{j_1}'(0)\,\L_{j_2}'(0)\,\L_{j_3}'(0) \cr
\eea
Notice that
\beq
\L_j'(0) = {(2j+1)!\over j! \,(j+1)!}\, (-2u_0)^{j+1} 2^{-2j}
\virg
\L_{-1}'(0)=1
\eeq

In particular this implies that:
\beq
{\partial^3 F_0\over \partial t^3} = {-1\over 2y'(0)} =  \dot u_0(t)
\eeq
and thus, as expected we recover:
\beq
\encadremath{
{\partial^2 F_0\over \partial t^2} =  u_0(t)
}\eeq

%\beq
%2 F_0 =  \Res_\infty \Phi_+\,\, y dx  
%\virg
%d\Phi=ydx
%\eeq

\subsubsection{Example $W_1^{(1)}$}

The recursion relation \eq{defrecursivspinvomng} gives:
\bea
{\cal W}_1^{(1)}(z) =W_1^{(1)}(z) x'(z)
&=& {1\over 8(\tau_3-2)}\,\left({1\over z^4}-{\tau_5\over (\tau_3-2)\, z^2}\right)  \cr
&=& {-\dot u_0\, \over 8}\,\left({1\over z^4}-{\dot u_0\over 3 z^2}\, \sum_l t_{l+1} (-u_0/2)^l \,{ (2l+3)!\over l! (l+1)!}\right)  \cr
\eea
and thus:
\bea
{\partial F_1\over \partial t_{j}}
&=& \Res_\infty {\cal W}_1^{(1)}(z)\, \L_{j}(z) \cr
&=& {2\dot u_0\over 3!\,\,\, 16}\left(\L_j'''(0) -2 \dot u_0\,\L_j'(0)\, \sum_l t_{l+1} (-u_0/2)^l \,{ (2l+3)!\over l! (l+1)!}   \right) \cr
\eea
In particular:
\bea
{\partial F_1\over \partial t}
&=& -{\dot u_0^2\over 24}\left( y'''(0) +  \sum_l t_{l+1} (-u_0/2)^l \,{ (2l+3)!\over l! (l+1)!}   \right) \cr
\eea

\subsubsection{Other examples}

\bea
{\cal W}_{2}^{(1)}(z_1,z_2) 
&=& {1\over 4(\tau_3-2)^2\, z_1^6 \,z_2^6}\,\Big[ {5!\over 2!} (z_1^4 <\psi_2^2>+z_2^4<\psi_1^2>)+ 3!^2 z_1^2 z_2^2 <\psi_1 \psi_2> \cr
&& \quad + \td{\tau}_1 z_1^2 z_2^4 <\kappa_1 \psi_1> + \td{\tau}_1 z_1^4 z_2^2 <\kappa_1 \psi_2>
+ {1\over 2} \td{\tau}_1^2 z_1^4 z_2^4 <\kappa_1^2 > \cr
&& + \td{\tau}_2 z_1^4 z_2^4 <\kappa_2>
\Big]
\cr
&=& {1 \over 8(\tau_3-2)^4 z_1^6 z_2^6}\,
\Big[ (\tau_3-2)^2 (5 z_1^4 + 5 z_2^4 + 3 z_1^2 z_2^2) + 6 \tau_5^2 z_1^4 z_2^4 \cr
&& - (\tau_3-2)(6 \tau_5 z_1^4 z_2^2 + 6 \tau_5 z_1^2 z_2^4 + 5 \tau_7 z_1^4 z_2^4) \Big]
\eea

\bea
{\cal W}_{1}^{(2)}(z) &=& - { 1 \over 128 (2-\tau_3)^7 z^{10}} \Big[ 252\, \tau_5^4 z^8 + 12\, \tau_5^2 z^6 (2-\tau_3) (50\, \tau_7 z^2 + 21\, \tau_5) \cr
&& \quad + z^4 (2-\tau_3)^2 ( 252\, \tau_5^2 + 348\, \tau_5 \tau_7 z^2 + 145\, \tau_7^2 z^4 + 308\, \tau_5 \tau_9 z^4) \cr
&& \qquad + z^2 (2-\tau_3) (203\, \tau_5 +145\, z^2 \tau_7 + 105\, z^4 \tau_9 +105\, z^6 \tau_{11}) \cr
&& \qquad \quad + 105\, (2 -\tau_3)^4 \Big] .\cr
\eea

\beq
{\cal W}_{4}^{(0)}(z_1,z_2,z_3,z_4) = 12\,{1\over (\tau_3-2)^3\, z_1^2 z_2^2 z_3^2 z_4^2}\,
\Big(  (\tau_3-2) (z_1^{-2} + z_2^{-2} + z_3^{-2}+ z_4^{-2})   - \tau_5 \Big) 
\eeq
and so on ...

\subsubsection{Homogeneity relation}

Theorem 4.7 of \cite{EOFg} gives another relation which we can apply here: the homogeneity equation.
Let $\Phi(z)$ such that
\beq
\Phi'(z) = y(z) x'(z) = {\L_{-1}\over -2 \dot u_0} \,\, 2z   = \sum_{j=0}^m t_j \L_j'
\eeq
and thus:
\beq
\Phi = \sum_{j=0}^m t_j \L_j
\eeq
We have:
\bea
(2-2g-n)\, {\cal W}_n^{(g)}(z_1,\dots,z_n) 
&=& \Res_0 {\cal W}_{n+1}^{(g)}(z_1,\dots,z_n,z_{z+1})\, \Phi(z_{n+1})  \cr
&=& - \sum_{j=0}^m t_j\, {\partial  \over \partial t_j}\,{\cal W}_n^{(g)}(z_1,\dots,z_n)   \cr
\eea
In other words, ${\cal W}_n^{(g)}$ is homogeneous of degree $2-2g-n$.

%\section{Boundary operators}

%All boundary correlation functions are given by the boundary descendants of the symplectic invariants of \cite{}.
%For example we have:
%\beq
%H_{(1)}^{(0)}(z_1,z_2) = {1\over z_1-z_2}\,\, {y(z_1)+y(z_2)\over z_1+z_2}
%\eeq
%\bea
%&& H_{(2)}^{(0)}(z_1,z_2,z_3,z_4) \cr
%&=&  {
%H_{(1)}^{(0)}(z_1,z_2) H_{(1)}^{(0)}(z_3,z_4) - H_{(1)}^{(0)}(z_1,z_4) H_{(1)}^{(0)}(z_3,z_2)\over (z_1^2-z_3^2)\,(y(z_2)-y(z_4))}  \cr
%&=&  {
%{y_1+y_2\over z_1^2-z_2^2}\, {y_3+y_4\over z_3^2-z_4^2} - 
%{y_1+y_4\over z_1^2-z_4^2}\, {y_3+y_2\over z_3^2-z_2^2}
%\over (z_1^2-z_3^2)\,(y(z_2)-y(z_4))}  \cr
%&=&   {(y_1+y_2)(y_3+y_4)(x_1-x_4)(x_3-x_2) - (x_1-x_2)(x_3-x_4)(y_1+y_4)(y_3+y_2)
%\over (x_1-x_2)(x_3-x_4)(x_1-x_4)(x_3-x_2)(x_1-x_3)\,(y_2-y_4)}  \cr
%&=&   {
%(y_1 y_3+y_2y_4)(x_1-x_3)(x_4-x_2)
%+(y_1y_4+y_2y_3)(x_1-x_4)(x_3-x_2)
%-(y_1y_2+y_3y_4)(x_1-x_2)(x_3-x_4)
%\over (x_1-x_2)(x_3-x_4)(x_1-x_4)(x_3-x_2)(x_1-x_3)\,(y_2-y_4)}  \cr
%&=& -{y_{13}\over x_{13}}\,x_{12}x_{34} - {x_{24}\over y_{24}}\,(y_1+y_2)(y_3+y_4)
%\eea

%\section{general $(p,q)$ model}

%\beq
%\left\{
%\begin{array}{l}
%x(z) = \sum_{k=0}^q u_k\, z^{q-k} \cr
%y(z) = \sum_{k=0}^p v_k\, z^{p-k} \cr
%\end{array}
%\right.
%\virg u_0=v_0=1\, ,\,\,\, u_1=v_1=0
%\eeq
%with:
%\beq
%\{y,x\}=1
%\eeq

\section{The $(2m,1)$ minimal model}

Another kind of universal limit of matrix models may arise when two connected components of the eigenvalues support merge, typically the equilibrium density of eigenvalues behaves as:
\beq
\hat y \sim x^{2m}.
\eeq
The case $m=1$ was treated in \cite{BlIt, BI2, BDJT}
The results concerning general $m$, were described without proof in \cite{BlEycrit}.
The universal limit is given by the $(2m,1)$ reduction of the mKdV hierarchy.
All the results can be proven in a way very similar to the $(2m+1,2)$ case, and here we just summarize the results the results stated in \cite{BlEycrit}.

\figureframex{14}{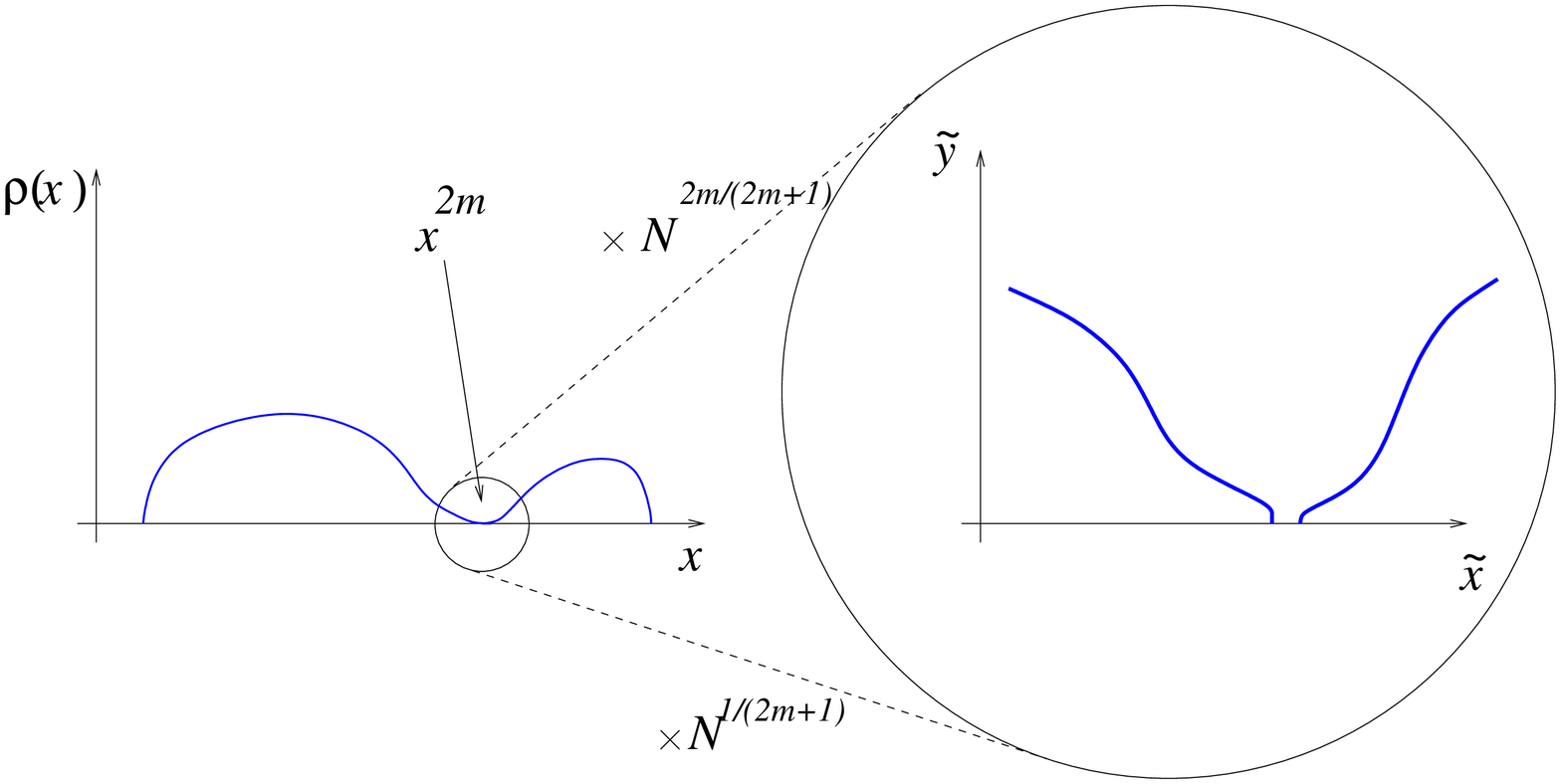}
{When two cuts merge,  the equilibrium density of eigenvalues behaves like $x^{2m}$ near the merging endpoints. The universal eigenvalues statistics is given by determinants of the $(2m,1)$ kernel, associated to the integrable mKdV hierarchy. \label{figdensitylim2m}}

\subsection{The $(2m,1)$ minimal model}

Again, we shall write a Lax pair ${\cal D}(x,t)$ and ${\cal R}(x,t)$ where ${\cal D}(x,t)$ is polynomial in $x$ of degree $2m$, and ${\cal R}(x,t)$ is polynomial of degree $1$, and such that:
\beq
 \left[{1\over N}  {\partial \over \partial x}-{\cal D}(x,t),{\cal R}(x,t)-{1\over N}\,{\partial \over \partial t}\right] = 0
\eeq

We choose:
\beq
{\cal R}(x,t) = \pmatrix{ 0 & x+u(t) \cr -x+u(t) & 0}
\eeq
and the matrix ${\cal D}(x,t)$ is of the form:
\beq
{\cal D}(x,t) = \sum_k t_k {\cal D}_k(x,t)
\eeq
with:
\beq
{\cal D}_k(x,t) = \pmatrix{-A_k(x,t) & x B_k(x,t)+C_k(x,t) \cr x B_k(x,t)-C_k(x,t) & A_k(x,t)}
\eeq
where $A_k$, $B_k$, $C_k$ are even polynomials of $x$, of degree $\deg A_k=2k-2$, $\deg B_k=2k-2$, $\deg C_k=2k$.
They can be found by recursion:
\bea
A_0 &=& 0 \virg B_0= 0 \virg C_0 = 1,\cr
C_{k+1} &=& x^2 C_k+\check R_k(u) \cr
B_{k+1} &=& x^2 B_k+\hat R_k(u) \cr
A_{k+1} &=& x^2 A_k+{1\over 2}\dot{\hat R_k}(u) \cr
\eea
where $\hat R_k(u)$ and $\check R_k(u)$ are the modified Gelfand-Dikii differential polynomials:
\bea
\hat R_0(u) &=& u \virg \check R_0(u)={u^2\over 2} \cr
\hat R_{k+1}(u) &=& u \check R_k(u) - {1\over 4} \ddot{\hat R}_k(u) \cr
\dot{\check R}_{k}(u) &=& u \dot{\hat R}_k(u) \cr
\eea
For example:
\bea
\hat R_1(u) = {u^3\over 2}-{\ddot u\over 4} &,& \check R_1(u) = {3 u^4\over 8}-{u \ddot u\over 4} + {\dot u^2 \over 8} \cr
\hat R_2(u) = {3 u^5\over 8}-{5 u^2 \ddot u\over 8} - {5 u \dot u^2 \over 8} 
&,&  
\check R_2(u) = {5 u^6\over 16}-{5 u^3 \ddot u\over 8}-{5 u^2 \dot u^2\over 16} -{u u^{(4)}\over 16}-{1\over 16}\,{\dot u {{\mathop{u}^{\dots}}}}\cr
&& \qquad \qquad  + {\ddot u^2 \over 32}  
\eea

The matrix ${\cal D}(x,t)$ satisfies
\beq
[\partial_x -{\cal D}(x,t), \partial_t-{\cal R}(x,t)]=0
\eeq
if and only if $u(t)$ satisfies the string equation:
\beq
\encadremath{
\sum_{k=0}^m t_k \hat R_k(u) = -t\, u.
}\eeq

The Baker-Akhiezer functions
\beq
\Psi(x,t) = \pmatrix{ \psi(x,t) & \phi(x,t) \cr \td\psi(x,t) & \td\phi(x,t)},
\eeq
are given by the common solutions of the two compatible systems:
\beq
{1\over N}\,{\partial \over \partial x}\,\Psi(x,t) = {\cal D}(x,t)\,\Psi(x,t)
\virg
{1\over N}\,{\partial \over \partial t}\,\Psi(x,t) = {\cal R}(x,t)\,\Psi(x,t).
\eeq
It was claimed in \cite{BlEycrit} that the parametrics asymptotics of orthogonal polynomials near the singularity are given by:
\bea
\psi_n((s-s_c)^{1\over 2m} x) 
&\sim& \cos{((n+{1\over 2})\pi\epsilon)}\, \psi(x) - \sin{((n+{1\over 2})\pi\epsilon)}\, \td\psi(x) \cr
&& + (s-s_c)^{1\over 2m} {u(t)\,\cos \pi\epsilon\over 4 (\sin\pi\epsilon)^2}\, \Big(\cos{(3(n+{1\over 2})\pi\epsilon)}\, \psi(x) \cr
&& \qquad - \sin{(3(n+{1\over 2})\pi\epsilon)}\, \td\psi(x)\Big) +\dots \cr
\eea
and where $\epsilon/(1-\epsilon)$ is the ratio of the number of eigenvales in the 2 cuts which merge at the singularity ($\epsilon=1/2$ is the symmetric case). 

\bigskip

The Christoffel-Darboux  kernel $K(x_1,x_2)$ is the same as \eq{detKCDD}:
\beq
K(x_1,x_2) = {\psi(x_1)\td\phi(x_2)-\td\psi(x_1)\phi(x_2)\over x_1-x_2}
\eeq
and the correlators are obtained by the same determinantal formulae \eq{detKW}.

And again the claim is that the determinantal correlators of the $(2m,1)$ minimal model, are the limits of matrix models correlators.

\section{Conclusion}

In this article, we have summarized some properties of scaling limits of matrix models (formal or not), known for a long time. We have provided a mathematical proof that the asymptotics of the $\hat \om_n^{(g)}$'s and $\hat f_g$'s are indeed those obtained from conformal field theory.
Our proof is based on the fact that the limits $\om_n^{(g)}$'s of matrix models correlators, are the spectral invariants of the limit spectral curve, and the fact that the determinantal correlators of the $(2m+1,2)$ minimal model kernel are also the spectral invariants of the same spectral curve.

\medskip

We recall that those correlators can be interpreted in the Kontsevich integral's framework, and have a combinatorial interpretation as intersection numbers of some tautological classes on the moduli spaces of Riemann surfaces.

\medskip

We claim that the same method can be applied to other sorts of universal limits, in particular the merging of two cuts like in \cite{BlEycrit}, and hopefully, we can work out the same kind of proof for multi-matrix models, whose universal limits should be the $(p,q)$ minimal models with arbitrary $p$ and $q$.
Unfortunately, one of the key points should be the equivalent of \cite{BergEyntopdet} (i.e. the fact that determinantal correlators obey loop equations), but this is not proved yet for differential systems of order $q>2$.

\section*{Acknowledgments}
We would like to thank A. Belavin, N. Orantin, J. Teschner, C. Tracy, S. Venakides, for useful and fruitful discussions on this subject.
This work is partly supported by the 
ANR project Grandes Matrices Al\'eatoires ANR-08-BLAN-0311-01, the European Science 
Foundation through the Misgam program, and the Quebec government with the FQRNT.

\end{document}